\documentclass[11pt,a4paper]{article}

\usepackage{jheppub}
\usepackage{epsfig}
\usepackage{latexsym}
\usepackage{amsfonts}
\usepackage{amsmath}
\usepackage{amsthm}
\usepackage{amssymb}
\usepackage{amsbsy}
\usepackage{multirow}
\usepackage{braket}
\usepackage{tikz}
\usepackage[boxsize=0.5em,centertableaux]{ytableau}
\usetikzlibrary{matrix,arrows}

\newcommand{\rar}{\rightarrow}

\newcommand{\tr}{\mbox{tr}}

\newcommand{\CC}{\mathbb{C}} 
\newcommand{\RR}{\mathbb{R}} 
\newcommand{\ZZ}{\mathbb{Z}} 

\newcommand{\h}{\mathfrak{h}}

\def\caln         {{\cal N}}
\def\calo         {{\cal O}}

\newsavebox{\uuunit}
\sbox{\uuunit}
    {\setlength{\unitlength}{0.825em}
     \begin{picture}(0.6,0.7)
        \thinlines
        \put(0,0){\line(1,0){0.5}}
        \put(0.15,0){\line(0,1){0.7}}
        \put(0.35,0){\line(0,1){0.8}}
       \multiput(0.3,0.8)(-0.04,-0.02){12}{\rule{0.5pt}{0.5pt}}
     \end {picture}}

\def\be{\begin{equation}}
\def\ee{\end{equation}}
\def\bea{\begin{eqnarray}}
\def\eea{\end{eqnarray}}

\def\a{\alpha}
\def\b{\beta}
\def\h{\eta}
\def\g{\gamma}

\def\d{\delta}
\def\e{\epsilon}
\def\D{\Delta}
\def\l{\lambda}
\def\L{\Lambda}

\def\f{\phi}

\def\m{\mu}

\def\o{\omega}

\def\p{\pi}
\def\r{\rho}
\def\z{\zeta}

\def\t{\tau}

\def\sF{{{ F}\!\!\!\!\hskip.8pt\hbox{\raise1pt\hbox{/}}\,}}
\def\som{{{ \omega}\!\!\!\!\hskip.8pt\hbox{\raise1pt\hbox{/}}\,}}
\def\sJ{{{\rm J}\!\!\!\!\hskip.8pt\hbox{\raise1pt\hbox{/}}\,}}

\def\pa{\partial}

\def\to{\rightarrow}
\def\nonu{\nonumber \\{}}
\def\half{{1 \over 2}}

\def\hsl{hs[$\l$]}
\def\vs{\vskip .1 in}

\def\Ab{\overline{A}}
\def\rar{\rightarrow}

\def\zb{\overline{z}}

\title{The semiclassical limit of $W_N$ CFTs and Vasiliev theory}
\author[a]{Eric Perlmutter,}
\author[b]{Tom\'{a}\v{s} Proch\'{a}zka}
\author[b]{and Joris Raeymaekers}
\affiliation[a]{DAMTP, Centre for Mathematical Sciences, University of Cambridge, CB3 0WA, UK}
\affiliation[b]{Institute of Physics of the ASCR,
Na Slovance 2, 182 21 Prague 8, Czech Republic.}
\emailAdd{E.Perlmutter@damtp.cam.ac.uk}
\emailAdd{prochazkat@fzu.cz}
\emailAdd{joris@fzu.cz}

\abstract{We propose a refinement of the Gaberdiel-Gopakumar duality conjecture between $W_N$ conformal field theories and 2+1-dimensional higher spin gravity. We make an identification of generic representations of the $W_N$ CFT in the semiclassical limit with bulk configurations. By studying the spectrum of the semiclassical limit of the $W_N$ theories and mapping to solutions of Euclidean Vasiliev gravity at $\lambda=-N$, we propose that the `light states' of the $W_N$ minimal models in the 't Hooft limit map not to the conical defects of the Vasiliev theory, but rather to bound states of perturbative scalar fields with these defects.  Evidence for this identification comes from comparing charges and from holographic relations between CFT null states   and bulk  symmetries. We also make progress in understanding the coupling of scalar matter to $sl(N)$ gauge fields.}

\begin{document}
\maketitle

\section{Introduction}

Holographic duality involving higher spin gravity appears to be a viable route toward understanding quantum gravity and the strong form of the AdS/CFT correspondence. Because computations can often be done on both sides at fixed values of the couplings, it is reasonable to seek fully soluble instances of duality. In the case of 3+1-dimensional higher spin gravity, this has recently been realized for free boundary field theories \cite{Klebanov:2002ja, Maldacena:2011jn}; one would of course hope to say the same of an interacting field theory, and perhaps one with tangible connections to string theory. There is already promise that this can be done in the AdS$_4$/CFT$_3$ context \cite{Giombi:2011kc, Chang:2012kt}, and that higher spin gravity itself can be extended beyond the current paradigm.

To this end, there has been recent movement toward understanding the space of conformal field theories with at least an approximate higher spin symmetry (in the sense of \cite{Maldacena:2012sf}) . In 1+1 dimensions, where the conformal symmetry is infinite-dimensional and interacting field theory realizations abound, this may be an especially rich pursuit. The most well understood example of this kind plays a central role in the specific proposal of Gaberdiel and Gopakumar \cite{Gaberdiel:2010pz, Gaberdiel:2012ku} -- on which we will focus in this paper -- which conjectures a duality between a 't Hooft limit of a certain class of minimal model CFTs and the 2+1-dimensional higher spin gravity theory of Vasiliev and collaborators. This field theory is (generically) interacting and has higher spin symmetry; its elegance derives in part from the fact that it possesses one current at each integer spin $s\geq 2$, and as such contains the minimal (albeit infinite-dimensional) field content needed to make sense of a possible higher spin duality.

Specifically, \cite{Gaberdiel:2010pz} proposed to study the $W_N$ minimal models, which can be realized as the coset
\be\label{cosss}
{SU(N)_k\otimes SU(N)_{1}\over SU(N)_{k+1}}
\ee
These theories are unitary generalizations of the Virasoro minimal models ($N=2$), with central charge
\be\label{cc1}
c = (N-1) \left(1 - {N(N+1) \over (N+k)(N+k+1)} \right).
\ee
Taking a 't Hooft limit in which $N,k\rar\infty$ with 't Hooft coupling $\l=N/(N+k)$ held fixed, the conjecture is that the limit theory is dual to 2+1-dimensional Vasiliev gravity based on gauge algebra \hsl\ with a single complex scalar field \cite{Prokushkin:1998bq}. Its backbone has been supported by a diversity of computations. Among the quantities that have been matched between bulk and boundary in the 't Hooft limit are the partition functions \cite{Gaberdiel:2010ar,Gaberdiel:2011zw}; three-point correlators between scalars and higher spin currents \cite{Chang:2011mz, Ahn:2011by, Ammon:2011ua}; a generalized Cardy formula to include higher spin charges in the presence of a spin-3 chemical potential \cite{Kraus:2011ds,Gaberdiel:2012yb}; and large $N$ factorization of correlators \cite{Papadodimas:2011pf,Chang:2011vka}.

Perhaps most convincingly of all, given the central role of symmetry in the AdS$_3$/CFT$_2$ correspondence, the asymptotic symmetry of the bulk theory \cite{Campoleoni:2010zq, Henneaux:2010xg, Gaberdiel:2011wb} known as $W_{\infty}[\l]$ has recently been proven to match the symmetry of the coset in the 't Hooft limit \cite{Gaberdiel:2012ku}. This is due to a `triality' of the quantum $W_{\infty}[\mu]$ symmetry of the coset (\ref{cosss}) at finite $c$, which is an isomorphism of algebras for three distinct values of $\mu$ . (This can also be understood heuristically as a generalized level-rank duality of the coset.) For the values of $\mu$ relevant to the holographic duality, the triality tells us that the quantum symmetries of bulk and boundary agree, and hence the duality at finite $N$ and $k$ should make sense.

There is another large $c$ limit we can take on the CFT side -- following \cite{Gaberdiel:2012ku}, we call this the `semiclassical limit' -- in which one holds $N$ fixed and takes $k$ such that $c$ becomes large. The theory retains $W_N$ symmetry but becomes non-unitary: for example, conformal weights become negative. Given the discovery of the full quantum $W_{\infty}[\mu]$ algebra of \cite{Gaberdiel:2012ku} which exists at continuous values of $c$, one can track a given representation of the theory as one takes $c$ large in either fashion described above. The behavior of the representations in the semiclassical limit can distinguish between perturbative (charges of order $c^0$) and solitonic quanta (charges of order $c$); this provides an `analytic continuation in c' of the minimal model representations.

In 1+1-dimensional CFT, large central charge $c$ implies the existence of a classical gravity dual. The aforementioned triality implies that the dual theory to the semiclassical limit of the $W_N$ CFT is the Vasiliev theory at $\l=-N$, the Euclidean version of which, as we elaborate upon in what follows, is a Chern-Simons theory with gauge algebra $sl(N,\CC)$ coupled to matter consistently with higher spin gauge invariance.

These ideas were utilized in \cite{Gaberdiel:2012ku} to modify the original proposal \cite{Gaberdiel:2010pz} and explain the presence of the so-called `light states' of the 't Hooft limit of the minimal models, that is, a discretuum of states which have zero charge in the 't Hooft limit. Conical defect solutions of the $sl(N,\CC)$ theory carrying ${\cal{O}}(c)$ charges, constructed in \cite{Castro:2011iw}, were proposed to be identified with the representations corresponding to the light states, but now in the semiclassical, rather than 't Hooft, limit. This analytic continuation was argued to give an indirect bulk description of the light states, in a different region of parameter space. A review of this story is given in \cite{Gaberdiel:2012uj}.

Such an identification was based largely on matching the charges of conical defects with those of the light states. In this work we will modify this proposal by taking a closer look at the spectrum in the semiclassical limit, studying the symmetries of the conical defect solutions, and constructing bound state solutions of scalars and defects in the Vasiliev theory at $\l=-N$.

Our findings are based on the following two main points, where for conciseness we borrow the familiar notation for $W_N$ representations which we explain in section 2:
\vs
1. To leading order in $c$ the semiclassical limit, generic representations $(\L^+,\L^-)$ have charges independent of $\L^+$. To distinguish among representations with different $\L^+$, one must examine the subleading ${\cal{O}}(c^0)$ pieces of the charges.
\vs
2. There is a direct map between classical symmetries of a given bulk solution and null vectors of the corresponding dual CFT representation.
\vs

We propose that, in fact, the conical defects are dual not to the light states $(\L^-,\L^-)$, but to the states $(0,\L^-)$. A generic light state is dual to a bound state of perturbative scalar fields (where a single scalar is dual to (f,0) as in \cite{Gaberdiel:2012ku}) and the conical defect. For instance, the simplest light state (f,f) is the bound state of a perturbative scalar field with the conical defect dual to (0,f). This follows upon examining linearized scalar and higher spin fluctuations around the conical defect backgrounds, and mapping to the structure of null vectors in the representation (0,$\L^-)$. This perspective is also encouraged by the (conjectural) form of the CFT partition function in the semiclassical limit, which can be written as a sum over saddles $(0,\L^-)$ and their 1-loop fluctuations. By using counting arguments (rather than direct calculation), we compute the bulk 1-loop partition function and find agreement with the CFT, within the range of validity of our computation.

In order to understand the scalar fluctuations, we needed to answer the prior question of what the physical content is of the linearized matter equations in the Vasiliev theory at $\l=-N$. For generic, i.e. non-integer, $\l$, these equations are well understood and have been used to compute correlation functions and probe higher spin black holes. At $\l=\pm N$, extra constraints arise on any scalar solutions. We argue that these constraints do not render the equations meaningless. Placing a scalar in both global and Poincar\'e AdS, one finds that it is restricted to live in a finite-dimensional non-unitary representation of the $sl(N,\RR) \times \overline{sl(N,\RR)}$ isometry of AdS. For $N=2$, these results fall into the classification of bulk highest weight representations given in \cite{Balasubramanian:1998sn}. We suggest that this makes sense via holography: the dual $W_N$ CFTs are non-unitary. Scalar fluctuations around conical defects support the above picture.

In all, we make progress in both understanding this semiclassical duality on its own terms, and as it relates to subtle aspects of the duality proposal of Gaberdiel and Gopakumar. Our computations are largely on the level of comparing spectra, and one would hope that the duality is robust to more detailed inquiry. The results on scalars in the Vasiliev theory at $\l=\pm N$ will hopefully find application beyond the scope of this work.

The remainder of the paper is organized as follows. Section 2 introduces the duality anew, lays out the CFT spectrum and partition function in the semiclassical limit and sets the stage for a detailed investigation of our proposal. In section 3, we establish the proper symmetry algebra organizing non-perturbative representations $(\L^+,\L^-)$ and give expressions for null states  which we will later match to symmetries of conical surpluses. Section 4 moves into the bulk, where we demonstrate in AdS that Vasiliev theory at $\l=\pm N$ admits certain scalar solutions that furnish non-unitary representations of the background isometry algebra. In sections 5 and 6, we put the pieces together by matching bulk solutions to boundary representations and computing the 1-loop partition function in the bulk. Section 7 concludes with a discussion. A series of appendices follows with supplements to various computations in the main text; we wish to point out Appendix D in particular, which gives details on the CFT representations $(f,0)$ and $(0,f)$ on which we will focus frequently.

\section{Large $c$ limits of $W_N$ conformal field theories}

We first set the stage by giving a basic description, mostly in words, of the current status of the Gaberdiel-Gopakumar duality proposal between large $c$ limits of $W_N$ minimal models and Vasiliev gravity. This is a slightly expanded version of the explanation in the introduction. We then move onto the details.

\subsection{A brief history of the $W_N$ minimal model duality}
Consider a CFT with $W_N$ symmetry, generated by conserved currents of spin $s=2,3,\ldots N$, and central charge $c$. We can choose to parameterize $c$ as
\be\label{cc}
c = (N-1) \left(1 - {N(N+1) \over (N+k)(N+k+1)} \right).
\ee
This is the so-called ``minimal model parameterization'' of the central charge: at the special values of $c$ where $k$ is a positive integer, the $W_N$ algebra allows for unitary minimal representations, which are realized by coset CFTs of the form
\be
{SU(N)_k\otimes SU(N)_{1}\over SU(N)_{k+1}}
\ee
These are the $W_N$ minimal models, higher spin generalizations of the Virasoro minimal models which are simply the $N=2$ theories above, and the subject of the original duality proposal of \cite{Gaberdiel:2010pz}. We will provide mathematical details of this construction in the next section; suffice it to say for now that the spectrum of minimal highest weight representations of $W_N$ is labeled by two highest weights of $su(N)_k$ and $su(N)_{k+1}$, denoted by $(\L^+,\L^-)$, respectively.
\vs

$\bullet$~{\bf Duality 1.0}\vs

Specifically, \cite{Gaberdiel:2010pz} proposed to study a particular large $c$ limit of the $W_N$ minimal model, namely a large $N$ 't Hooft limit
\be
 N, k \rar \infty~,\quad    \lambda \equiv {N \over k+N}~~{ \rm fixed}
\ee
In the limit, the central charge scales like
\be
c\sim N(1-\l^2)
\ee
with $0\leq \l\leq 1$ and the theory remains unitary. This theory was conjectured to be dual to Vasiliev's theory of higher spin gravity, with gauge fields valued in the Lie algebra hs$[\l]$ coupled nonlinearly to two complex scalar fields each of mass $m^2=-1+\l^2$. (For some details on \hsl, see references \cite{Gaberdiel:2011wb, Ammon:2011ua}, for instance.)

The scalar fields were conjectured to be dual, in the 't Hooft limit, to the primaries (f,0) and (0,f) and their complex conjugates, where `f' denotes the fundamental representation. These primaries have conformal dimensions $\Delta_+=1+\l$ and $\Delta_-=1-\l$, respectively, so the bulk scalars are quantized appropriately; indeed, the scalar mass is in the window for alternative quantization to be an option. The (quantum) symmetries of bulk and boundary were recently proven to match: they are given by the infinite-dimensional algebra $W_{\infty}[\mu]$, the quantum version of which has a triality under which distinct algebras are isomorphic. For $\mu=\l$, the classical version of which is the asymptotic symmetry of the bulk theory \cite{Gaberdiel:2011wb}, the algebra is isomorphic to the algebra at $\mu=N$, labeled $W_{N,k}$, which is that of the boundary CFT:
\be
W_{N,k} \cong W_{\infty}[\l]
\ee
Because we are dealing with the quantum algebra, the isomorphism is valid for finite $N$ and $k$, including the 't Hooft regime in which both become large; this then explains the match of classical symmetries. Much more can be said about this proposal, and we point the reader to the recent, detailed review \cite{Gaberdiel:2012uj}.

One uncomfortable aspect of this proposal was the existence of so-called `light states' in the CFT, representations in which $\Lambda^+ = \L^-$. These have vanishing charges and form a discretum in the 't Hooft limit, and decouple from all correlation functions strictly in the limit; however, the duality cannot account for them at finite $N$, as the states are no longer decoupled, and the bulk Vasiliev theory does not have the required degrees of freedom \cite{Gaberdiel:2011zw, Papadodimas:2011pf, Chang:2011mz} .\vs

$\bullet$~{\bf Duality 2.0}\vs

The latest incarnation of this conjecture, put forth in \cite{Gaberdiel:2012ku}, modifies the above picture. Given our knowledge of the full quantum $W_{\infty}[\mu]$ algebra, one can take a diferent large $c$ limit -- the `semiclassical' limit -- in which $N$ is held fixed and finite while $c$ is taken to infinity. With the parameterization (\ref{cc}), note that for a given value of $c$ there are two branches for $k$. These are related by triality of \cite{Gaberdiel:2012ku}, and in what follows we will choose the branch
\be k = - N-\frac{1}{2} -\frac{1}{2} \sqrt{1-\frac{4 N \left(N^2-1\right)}{c-N+1}}.\ee
for which, in the limit of large $c$ for fixed $N$, $k$ goes like
\be\label{klargec}
k\rar -N-1+{N(N^2-1)\over c}+\calo( 1/c^2)
\ee

In the semiclassical limit, the representations of the algebra become non-unitary (e.g. conformal dimensions become negative). These can be thought of as analytic continuations in $c$ of the corresponding minimal model representations. The behavior of these representations in the semiclassical limit can distinguish between perturbative and solitonic quanta. That is to say, the two distinct classical limits of the quantum algebra may {\it a priori} have different spectra. In particular, one finds that representations with $\L^-=0$ become light in the semiclassical limit -- with charges of $\calo(c^0)$ -- and those with $\L^-\neq 0$ become heavy, with charges of $\calo(c)$. This latter group includes both the (0,f) representation and the light states, which are no longer light in the semiclassical limit.

Despite the non-unitarity of the semiclassical limit of the minimal models, large $c$ implies the existence of a gravity dual. In particular, the aforementioned triality tells us that the Vasiliev theory with $\lambda=-N$ has the correct symmetry. At $\l=-N$, an ideal of \hsl\ forms, and the Lie algebra becomes $sl(N)$ upon factoring out this ideal: thus, the gravity dual is the $sl(N)$ Vasiliev theory\footnote{The Lorentzian theory has two copies of $sl(N,\RR)$ as its symmetry algebra, while the Euclidean theory has one $sl(N,\CC)$ algebra. When speaking in generalities, we will often refer simply to $sl(N)$.}, describing a tower of higher spin fields $s=2,3,\ldots N$ and its nonlinear, gauge invariant coupling to matter.

The above discussion implies that representations $(\L^+,\L^-)$ with $\L^-\neq 0$ should be identified, in the semiclassical limit, with classical backgrounds of the $sl(N)$ theory, perhaps coupled to perturbative scalar quanta as dictated by $\L^+$. This identification then analytically continues back to the 't Hooft limit. In \cite{Castro:2011iw} a set of conical surplus solutions to the Euclidean $sl(N,\CC)$ gravity theory were argued to lie in one-to-one correspondence with the representations $(\Lambda^-,\Lambda^-)$ in the semiclassical limit, based on a nontrivial matching of charges to leading order in $c$.

In the final proposal of \cite{Gaberdiel:2012ku}, then, the dictionary for mapping CFT representations in the 't Hooft limit to bulk configurations was as follows:\vs

$\bullet$ The (f,0) representation, together with its complex conjugate ($\bar f$,0), should remain dual to a single complex scalar of the bulk \hsl\ theory, with standard quantization. \vs

$\bullet$ One then identifies the light states $(\L^-,\L^-)$ with the analytic continuation of the conical surplus solutions of the $sl(N)$ theory back to the unitary regime. Any other state $(\L^+,\L^-)$ with $\L^+\neq \L^-$ maps to the analytic continuation of a configuration of perturbative scalars propagating on a given conical surplus background.\vs

Strictly in the 't Hooft limit, the original duality proposal needed no modification because the light states decouple; it is the consideration of what happens at finite $N$ which compels this change. While the quantum higher spin gravity theory has yet to be constructed, the expectation is that the spectrum undergoes some non-perturbative truncation of all spins $s>N$, and the above map between defects and light states continues to hold.

It is this map between bulk solitons and CFT light states that we will modify in the present work. To do so, we must introduce the technology needed to describe representations of the $W_N$ algebra, and then understand the consequences of the large $c$ limit. The punchline is summarized in subsection 2.4.

\subsection{Maximally degenerate representations of $W_N$ algebra}\label{secdegreps}
The $W_N$ algebra \cite{Zamolodchikov:1985wn} (see \cite{Bilal:1991eu,Bouwknegt:1992wg} for reviews and further references) is a nonlinear  extension  of the Virasoro algebra generated
by the Virasoro generators  $L_n$ and the modes $W^s_n$ of the higher spin primaries, where $s = 2, \ldots , N$. Some of the
commutation relations are given in Appendix \ref{appWn}. The $W_N$ algebra depends on a single parameter, the central charge $c$, which can be varied continuously.
Even though we will mostly  discuss a single holomorphic $W_N$ algebra, we have in mind theories which also possess
an antiholomorphic $\overline{W}_N$ symmetry with generators $\overline{ L}_n, \overline{ W}^s_n$, and all of our results apply equally to that sector.

The representations we will be interested in are the irreducible highest weight representations of $W_N$. The usual construction of these representations proceeds in two steps. First one constructs a Verma module with highest weight vector $\ket{w_s}$. The highest weight vector or primary  is annihilated by all positive modes of $W_N$ and is an eigenvector of all zero modes with eigenvalues $w_s$. All other vectors of the Verma module are obtained by acting on $\ket{w_s}$ with negative modes using only the commutation relations of the $W_N$ algebra.

Depending on the choice of the central charge $c$ and $w_s$ charges, this representation may or may not be irreducible. The representation is reducible if it contains at least one
null state, i.e. a state which is both descendant and primary.
Such a null state is the highest weight of a Verma submodule whose states have vanishing BPZ inner product with all other states in the Verma module and can be consistently factored out. They represent a non-trivial relation between $W_N$ generators in the module that we can impose apart from the commutation relations. Factoring out all the Verma submodules leads us to an irreducible representation of $W_N$ algebra.

The representations appearing in the usual unitary $W_N$ minimal models are such that the corresponding Verma modules are maximally degenerate. The corresponding primaries $\ket{w_s}$ are labeled by two  representations $(\Lambda^+, \Lambda^-)$ of $su(N)_k$ and $su(N)_{k+1}$.
 For such representations, the  Verma modules contain an infinite number of Verma submodules. For this to be possible, the central charge $c$ must be equal to one of a discrete set of values bounded from above by $N-1$.

Here we instead want to study representations of $W_N$ at fixed $N$ and large (generic) $c$. This regime is most conveniently analyzed  using a free field realization \cite{Fateev:1987vh,Fateev:1987zh}. It turns out that even in this regime the primaries labeled by two $sl(N)$ representations $(\Lambda^+,\Lambda^-)$ are maximally degenerate, i.e. they contain the maximal number of null vectors possible (for the fixed generic value of $c$).  We will presently review these representations and their properties.

In the following we represent $sl(N)$ weight vectors as $N$-dimensional vectors whose components sum to zero. The inner product\footnote{We hope it is clear from the context whether $(\L^+,\L^-)$ stands for a representation or an inner product of weight vectors.} $( \cdot , \cdot)$ in weight space is the Euclidean inner product
 in $\RR^N$.  The highest weight corresponding to a Young tableau with $r_i$ boxes in the $i$-th row (and $r_N\equiv 0$) has components
 \be
\L_i = r_i - {B \over N}\qquad i = 1, \ldots , N\label{defLambda}
\ee
where $B$ is the number of boxes in the diagram. The Weyl vector $\r$ has components
 \be
\r_i = {N+1\over 2} - i\qquad i = 1, \ldots , N.\label{defWeyl}
\ee
It will also be useful to define the vectors $n^{\pm}$ as
\be
n^\pm = \L^\pm + \r.\label{defn}
\ee

All the $W$-charges of the highest weight state $\ket{\Lambda^+,\Lambda^-}$ are determined in terms of $\L^+$ and $\L^-$. In particular, the conformal weight $h \equiv w_2$ is equal to
\bea
h_{(\L^+,\L^-)} & = & \frac{1}{2(N+k)(N+k+1)} \left[ \left((N+k+1)(\Lambda^++\rho) - (N+k)(\Lambda^-+\rho)\right)^2 - \rho^2 \right] \nonu
  & = & C_2(\theta) + \frac{c-N+1}{24}\label{weight}
\eea
where we have introduced  the power sums $C_s (\theta) = \sum_i (\theta_i)^s/s$  and $\theta$ is the vector
\bea
\theta &=& \a_+n^+  + \a_- n^-\\
\a_+ &=& \sqrt{ N + k + 1 \over N+k} \qquad \a_- = - {1 \over \a_+}\label{thetaeq}
\eea
The spin 3 charge $w_3$ of the primary in the $(\L^+,\L^-)$ representation is given by
\be
w_{3,(\L^+,\L^-)}
  = \g C_3 (\theta)\label{spin3charge}
\ee
where the constant $\g$ in (\ref{spin3charge}) normalizes the spin 3 modes
 as in (\ref{wcomm}).\footnote{
The explicit expression is $\g = - \sqrt{ 2 N (N^2-1) \over (N-2)(3N^2 + (c-1)(N+2))}$.}  This implies that $w_s\propto\g^{s-2}$; in addition, $w_s$ is of $\calo(\theta^s)$.

\subsubsection{Structure and characters of degenerate representations}\label{secstruct}

Let us now look have a closer look at the structure of irreducible representations parametrized by $(\Lambda^+, \Lambda^-)$ at generic values of $c$. We have two goals: we need to understand what are the independent null states in the corresponding Verma module and we want a formula for the character. `Independent' null vectors have the property that all other null vectors are their descendants.

First let us consider the vacuum representation. It corresponds to taking $\Lambda^+ = \Lambda^- = 0$ with all highest weight charges equal to zero, $w_s = 0$. This primary is  annihilated not only by positive and zero modes of $W_N$ generators, but also by some of the negative modes,
\be
W^{s}_{-j} \ket{0} = 0 \quad \quad j = 1, \ldots, s-1 \label{cftvacinv}
\ee
This is an $N>2$ generalization of the global conformal invariance of the Virasoro vacuum state. As discussed in the previous section, these additional constraints in the irreducible module manifest themselves through the presence of null states in the $\ket{0}$ Verma module. In this example, one can show that there are no additional null states apart from those coming from (\ref{cftvacinv}).

The spectrum of states in a highest weight representation is captured by the character ${\rm ch} = {\rm Tr} q^{L_0-c/24} $.
The character of a representation with conformal weight $h$ which counts all the states in the Verma module  is
\be
{\rm ch}_h = {\rm Tr}_V q^{L_0-c/24} = q^{h-c/24} \prod_{j=1}^\infty \frac{1}{(1-q^j)^{N-1}}
\ee
where the trace runs over all Verma module states.
For degenerate representations, this must be corrected due to the null states. For the vacuum representation, it is easy to see from
(\ref{cftvacinv}) that this correction takes the form
\bea
{\rm ch}_{(0,0)} &=&   {\rm ch}_{h=0} \times \prod_{s=2}^N \prod_{j=1}^{s-1} (1-q^j)\label{polcorr}\\
&=& q^{-c/24} \prod_{s=2}^N \prod_{j=s}^\infty \frac{1}{1-q^j} \label{vacchar}
\eea

For general representations $(\Lambda^+,\Lambda^-)$, the correction factor in  (\ref{polcorr})  due to the null states is  replaced by a more complicated polynomial which encodes the structure of the $(\Lambda^+,\Lambda^-)$ Verma module. For the regime of interest (large and generic $c$)  this structure was worked out in \cite{Niedermaier:1991cu}, which we  review and motivate using examples in Appendix \ref{appVerma}.
One property that will play an important role in what follows is that every $(\Lambda^+,\Lambda^-)$ representation
contains  $N-1$  independent null vectors.
These appear at levels
\be
(\L^+_{j}-\L^+_{j+1}+1)(\L^-_{j}-\L^-_{j+1}+1) ~,\qquad j = 1, \ldots , N-1.\label{nulllevels}
\ee
For the vacuum representation, the independent null vectors all appear at level one and are given by $L_{-1} \ket{0}, W^3_{-1} \ket{0}, \ldots ,  W^N_{-1} \ket{0}$.\vskip .1 in

From these considerations, one can derive the character formula for general $(\L^+,\L^-)$ representations  in the generic large $c$ regime of interest \cite{Niedermaier:1991cu}. The
result is
\bea
{\rm ch}_{(\L^+,\L^-)} & = & q^{-c/24}\prod_{j=1}^\infty \frac{1}{(1-q^j)^{N-1}} \sum_{w \in W} \epsilon(w) q^{h(w \cdot \Lambda^+, \, \Lambda^-)} \nonu
& = & {1 \over \h^{N-1}} \sum_{w\in W} \e ( w) q^{ {1\over 2(N+k)(N+k+1)} \left( (N+k+1) w(\L^+ + \r) - (N+k) (\L^- + \r)\right)^2}\label{char}
\eea
where the sum runs over the Weyl group of the finite dimensional $sl(N)$ algebra\footnote{Note that for unitary minimal models -- where $c$ is given by (\ref{cc}) with $k$ a positive integer -- the character formula is different and it involves a sum over an affine Weyl group reflecting the fact that there are infinitely many additional null states.}.

Now we turn to the partition function capturing the full spectrum of the CFT. The $W_N$ minimal model CFTs which feature in the holographic duality proposal of \cite{Gaberdiel:2010pz} contain each maximally degenerate representation once, leading to the diagonal modular invariant partition function. In analogy to this, we will consider here  a conformal field theory which  contains each  of the
$(\L^+,\L^-)$ representations once, transforming in the same way under the holomorphic and antiholomorphic part of the symmetry algebra. This then constitutes our proposal for the continuation of the $W_N$ minimal model CFTs to the large $c$ regime.
The partition function is given by the diagonal sum of the characters
\bea
Z_{CFT} &=& {\rm Tr} q^{L_0 - c/24}  \bar q^{\bar L_0 - c/24}\nonu
&=& \sum_{\L^+, \L^-} |{\rm ch}_{(\L^+,\L^-)} |^2.
\eea

Let us rewrite this result for later convenience. First we observe that, using the Weyl character formula, the character (\ref{char}) can be
rewritten as
\be
{\rm ch}  _{(\L^+,\L^-)} = {\rm ch}_{(0,\L^-)} \  q^{{N+k+1\over N+k} \left( C_2( n^+ ) - {N(N^2-1)\over 24}\right) }\chi_{\L^+} ( - 2 \p i \t n^-)
\ee
where $\t$ is defined through  $q = e^{2 \p i \t}$ and $\chi_{\L} (h) $ is the finite $sl(N)$ character evaluated on the weight vector $h$.

We can write a more explicit expression for ${\rm ch}_{(0,\L^-)}$ using the Weyl
denominator formula:
\be
{\rm ch}_{(0,\L^-)} =  {q^{h_{(0,\L^-)} - {c- (N-1)\over 24}} \over \h^{N-1}} \prod_{1 \leq i < j \leq N} (1 - q^{n^-_i-n^-_j}).
\ee
The CFT partition function can then be rewritten as
\be\label{cftp}
Z_{CFT} = \sum_{\L^-} Z_{\L^-}
\ee
with
\be
Z_{\L^-} = |{\rm ch}_{(0,\L^-)}|^2 \sum_{\L^+} (q\bar q)^{{N+k+1\over N+k} \left( C_2( n^+ ) - {N(N^2-1)\over 24}\right)  } |\chi_{\L^+} ( - 2 \p i \t n^-)|^2.\label{CFTZ}
\ee

In the `semiclassical' large $c$ approximation where we keep the terms up to order 1 in the $q$-exponents, this expression reduces to
\be
Z_{\L^-} \sim {|q|^{ - {2 C_2(n^-) \over N(N^2-1)} (c -(N-1)) + 2 (\L^-,\L^- + \r)} \over |\h |^{2(N-1)}} \prod_{1 \leq i < j \leq N} |1 - q^{n^-_i-n^-_j}|^2 \sum_{\L^+} |\chi_{\L^+} ( - 2 \p i \t n^-)|^2.\label{CFTZlargec}
\ee
This is the expression we wish to  compare with the semiclassical partition function in the bulk dual higher spin gravity theory.

\subsection{Perturbative and nonperturbative states at large $c$}\label{seclargec}

Because the quantum $W_N$ algebra exists for continuous values of $c$, we can track its maximally degenerate representations in the large $c$ limit keeping $N$ fixed.
For large $c$, the conformal weights and higher spin charges have a semiclassical  expansion
\be
 w_{s,(\L^+,\L^-)}  = w^{(-1)}_{s,(\L^-)} c  + w^{(0)}_{s,(\L^+,\L^-)} + \calo (1/c).\label{chargeslargec}
 \ee
Looking back at (\ref{klargec}) and (\ref{thetaeq}), we see that $\alpha_+\sim c^{-1/2}$ and $\alpha_-\sim c^{1/2}$ in the large $c$ limit. This means that, for a given representation $(\Lambda^+,\Lambda^-)$, its charges $w_s$ are independent of $\Lambda^+$ to leading order in $c$, as indicated in the above expression.
We see this explicitly in the behavior of spin 2 and spin 3 charges (\ref{weight}), (\ref{spin3charge})  at large $c$:
\bea
h^{(-1)}_{(\L^-)}  &=& - {C_2 (n^-) \over N(N^2-1)} + {1 \over 24} \label{O1weight} \\
h^{(0)}_{(\L^+,\L^-)}  &=&  {C_2  (n^- )\over N(N+1)}-  {N-1\over 24}+(\Lambda^-,n^-)-(\Lambda^+,n^-) \label{O0weight}\\
w^{(-1)}_{3, (\L^-)}  &=& - \frac{ \sqrt{2}  }{ N  (N^2-1)\sqrt{4 - N^2}} C_3 (n^-) \label{O1spin3} \\
w^{(0)}_{3,(\L^+,\L^-)}  &=&\frac{
   \sqrt{2} \left(3 N^3+9 N^2+9 N+4\right) }{ N
   (N+1) (N+2) \sqrt{4-N^2}}C_3 (n^-) - \frac{ \sqrt{2} }{\sqrt{4-N^2}} \sum_i (n^-_i)^2 n^+_i
. \label{O0spin3}
\eea
where we recall the definition $n^\pm = \L^\pm + \r$.

As expected, the maximally degenerate representations have negative conformal weights  in the large $c$ limit (except for the vacuum representation) and are hence  non-unitary.
We also see from these expressions that the maximally degenerate representations fall into two classes with dramatically different large $c$ behavior, which will correspondingly have rather different bulk interpretations as discussed in \cite{Gaberdiel:2012ku}. Representations of the type $(\L^+, 0)$ have (negative) energies  of order one. In the standard AdS/CFT dictionary, such states are expected to correspond to perturbative single- and multiparticle states in the bulk
theory expanded around the AdS background. On the other hand, the $(\L^+, \L^-)$ representations with $\L^-$ nonzero have large (negative)
  energies of order $c$. These are not expected  to correspond to perturbative bulk states: rather, they will correspond to classical gravity backgrounds, perhaps with perturbative excitations as dictated by $\Lambda^+$.

\begin{figure}[t]
\begin{center}
\begin{picture}(400,100)
\put(50,0){\includegraphics[width=250pt]{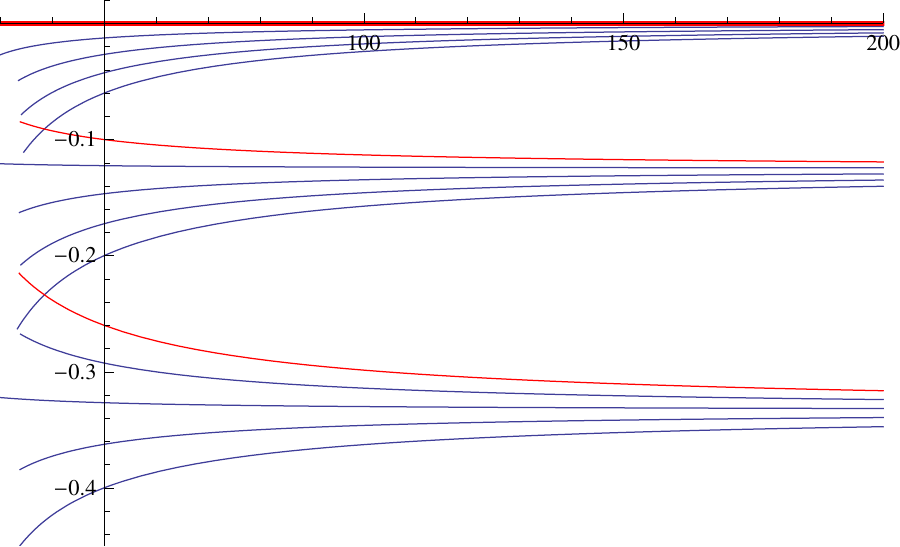}}
\put(290,152){$c$}
\put(60,-5){$h/c$}
\end{picture}
\end{center}
\caption{Large $c$ behavior of Virasoro ($N=2$) maximally degenerate representations. The plot shows $h/c$ as a function of $c$. The red lines (top in each band)
are the first  three $(0, \L^-)$ representations (corresponding to Young diagrams with $0,1,2$ boxes), while the blue lines  denote  the  states $(\L^+, \L^-)$ for the first four  representations $\L^+$ (Young diagrams with $0,1,2,3$ boxes). }\label{Nis2largec}
\end{figure}

Rewriting (\ref{O1weight}), (\ref{O0weight}) as
  \be
h_{(\L^+,\L^-)} = h_{(0,\L^-)} - (\L^+, n^-) + \calo(1/c)\label{endiff}
\ee
and observing that $(\L^+,  n^-)$ is always positive\footnote{This can be seen as follows: $\L^+$ and $ n^-$ have positive coefficients when expanded in the basis of fundamental weights, and in $sl(N)$ the inner product between any two fundamental weights is positive.}, we see that
within a sector of fixed $\L^-$ the primary $(0, \L^-)$ has the highest energy, while turning on $\L^+$ produces a `band' of closely spaced states with lower energy. The same goes for all higher spin charges. This is
 illustrated for the Virasoro ($N=2$) maximally degenerate representations in figure \ref{Nis2largec}.

Another property which sets the $(0, \L^-)$ primaries apart from the other $(\L^+, \L^-) $  states is the behavior of their independent null states
at large $c$. This is already evident in the simplest example of $N=2$: from (\ref{nulllevels}), a general $(\L^+, \L^-) $ primary has a null state at level $(r^+_1+1)(r^-_1+1)$
where $r^\pm_1$ are the numbers of boxes in the Young diagrams of $\L^\pm$.\footnote{ The standard notation (see e.g. \cite{DiFrancesco:1997nk}) for the degenerate Virasoro representations uses the labels $(r,s)$ instead of our $r^\pm_1$. These are related  as $r = r^-_1 + 1, s = r^+_1 +1$.}
The large $c$ behaviour of the null state is   \cite{Bauer:1991ai}:
\be
\left(\left(L_{-(r^-_1 + 1)}\right)^{r^+_1 + 1} + \calo (1/c) \right) \ket{\L^+, \L^-}\label{nullNis2}
\ee
From (\ref{nullNis2}) we see that this leading operator is linear  only for the $(0, \L^-) $ primaries.
 This property  appears   to hold for general $N$;  for example,  from  (\ref{fnullvects}) we see that
 the independent null vectors of the $(0,f)$   representation linearize for general $N$, while those of  the $(f,0)$ representation do not. We should note that, since the states $\ket{\L^+, \L^-}$ have quantum numbers of order $c$ in general, the neglected $\calo(1/c)$ part is crucial to show that the state is indeed annihilated by all positive modes. Nevertheless, we shall see that the leading part of the operator producing the
null state displayed above has a clear counterpart in the bulk theory.

Further examples of this linearization property for the $N=3$ case are given in table \ref{nulltab} of Appendix \ref{appnull}.
 We conjecture that this property of the $(0,\L^- )$ independent null states  holds  for general $N$.
  It  will be shown to reflect the fact that the associated
bulk classical solution has additional symmetries.

\subsection{A modified duality proposal}\label{secmoddual}

Having laid out some aspects of these representations in large $c$ limit, we can address the main question at hand: which CFT representations map to which bulk configurations? In contrast to the proposal of \cite{Gaberdiel:2012ku} briefly outlined in subsection 2.1.2, we now make the following concrete conjecture. The $(0,\L^-)$ states, and not the $(\L^-,\L^-)$ states, correspond to the smooth conical solutions constructed in \cite{Castro:2011iw}. The energy of a general primary (\ref{weight}) can be written as
\be
h_{(\L^+,\L^-)} = h_{(\L^+,0)} + h_{(0,\L^-)}  - (\L^+, \L^-)
\ee
suggesting that a general primary $(\L^+,\L^-)$ can be seen as a bound state of the solitonic solution $(0,\L^-)$  and the perturbative  excitation $(\L^+, 0)$; similar statements can be made about all higher spin charges.

This will be borne out by our bulk analysis. First, we will map the null states of the $(0,\L^- )$ representations to symmetries of the surplus backgrounds. This is much in the spirit of \cite{Maloney:2007ud,Castro:2011ui,Castro:2011zq} and, more generally, the role of gauge invariance in three-dimensional gravity. We will also map the $\calo(c^0)$ pieces of the $(\L^+,\L^-)$ charges to zero mode eigenvalues of scalar fluctuations: as has been understood in the context of pure AdS$_3$ gravity \cite{Balasubramanian:1998sn}, repeated action of bulk differential operators on solutions to the equations of motion generates entire representations of the background isometry algebra.

Our proposal is also suggested by the form of the CFT partition function (\ref{cftp}), which has the form of a sum over saddle points $(0,\L^-)$ and the fluctuations around them. It will be our goal below to explain these expressions from the bulk Vasiliev theory: we will argue that the sum over $\L^-$ corresponds to a sum over classical saddles -- the conical surplus solutions -- and the inclusion of the 1-loop determinant
of scalar and higher spin fluctuations around each saddle will nearly reproduce the full 1-loop piece of $Z_{\L^-}$ (\ref{CFTZlargec}).
These arguments are indirect -- we have not been entirely rigorous in showing that (\ref{cftp}) is indeed the large $c$ partition function, nor will we derive it in the bulk directly -- but suggestive nonetheless.

  For later reference, we record here the differences -- at large $c$ -- in energy and spin 3 charge between the primaries $(f,\L^-)$ and $(0,\L^-)$, which can easily
  be obtained from (\ref{O1weight}), (\ref{O1spin3}):
\bea
h_{(f,\L^-)} - h_{(0,\L^-)} &=& - n^-_1  + \calo ({1/ c}).\label{fLambdaweight}\\
w_{3,(f,\L^-)} - w_{3,(0,\L^-)} &=& i\sqrt{2 \over N^2-4} \left((n^-_1)^2 - {1\over N} (n^-,n^-)  \right) + \calo ({1/ c}).\label{fLambdaspin3}
\eea

Before moving into the bulk, we establish the proper symmetry algebra which acts on the nonperturbative states.

\section{Structure of $W_N$ representations at large $c$}\label{sectwisted}
As was shown by Bowcock and Watts \cite{Bowcock:1991zk}, the $W_N$ algebras  contain,  in the large  $c$ limit, an  $sl(N)$ subalgebra which organizes the spectrum of perturbative states: each perturbative primary corresponds to an $sl(N)$ highest weight and vice versa.
This is the algebra generated by the modes $W^s_m$ in the `wedge' $|m|<s$ which annihilate the vacuum. Indeed, for these modes the central terms in their commutators vanish, while the nonlinear terms are suppressed at large $c$. We will illustrate this in more detail below.

For the nonperturbative states, which have higher spin  charges that are proportional to $c$, the zero mode generators should be viewed as being  `of order $c$' and the argument for the
suppression of nonlinear terms in the commutators of the wedge modes fails. Nevertheless, we shall now argue that even for the nonperturbative states there is a set of generators which form an approximate $sl(N)$ algebra at large $c$  and which organize the spectrum of nonpertubative states. We will refer to these generators as `twisted wedge' generators.

First we introduce redefined generators which remain finite at large $c$ when acting on a class of nonperturbative states. We fix a representation $\L^-$ and define renormalized generators ${W^s_m}'$ with shifted zero modes:
\be
{W^{s}_m}' = W^s_m - w_{s ,(0,\L^-)} \d_{m,0}.
\ee
Since the charges of all primaries $(\L^+,\L^-)$ and their descendants have the same leading part as the  $(0,\L^-)$ primary, the new zero modes are finite acting on all states with $\L^-\neq 0$. We have chosen the order one part of the renormalization such that the primary  $(0,\L^-)$ now has vanishing charges, anticipating that $(0,\L^-)$ will play the role of the vacuum in the nonperturbative sector labeled by $\L^-$.

In terms of the redefined generators, suppressing the subscripts $\,_{(0,\L^-)}$ for brevity, the first few commutation relations (\ref{wcomm}) become
 \begin{align}\label{wprimecomm}
{}[L'_m, L'_n] &= (m-n)L'_{m+n} +\left( { c\over 12} m(m^2-1) + 2m  h \right)  \delta_{m,-n} \nonumber\\ \
{}[L'_m, {W'}^s_n] &= \left((s-1)m-n\right){W'}^s_{m+n}+ s mw_{s } \d_{m,-n}\nonumber \\
{}[{W'}^3_m, {W'}^3_n] &= 2(m-n){W'}^4_{m+n}+  (m-n)\left(\frac{1}{30}(2m^2+2n^2-mn-8) + {32\over 5 c  + 22}h \right)L'_{m+n} \nonumber \\
 & \hspace{-60pt} +\frac{16 }{(5 c+22 )}(m-n){\Lambda^{(4)}}'_{m+n} \nonumber \\
  &  \hspace{-60pt} + \left(\frac{c }{3 \cdot 5!}m(m^2-1)(m^2-4) + 4 m w_{4 } + {32\over 5 c +22} m (
 h^2 + {1 \over 5} h) + {1 \over 15} m(5 m^2 -8)h \right)  \delta_{m,-n}\nonumber \\
\end{align}
We now take the large $c$ limit appropriate for comparing with the classical theory in the bulk as advocated in \cite{Bowcock:1991zk}.
 This proceeds in two steps. We recall that, in terms of bulk quantities $c \sim {l \over \hbar G}$ where $l$ is the AdS radius
 and $G$ is Newton's constant.
 First we `restore $\hbar$' by rescaling
\bea
 {\tilde W^{s}_m} & = & \hbar {W^{s}_m}'\\
 \tilde c & = & \hbar c.
\eea
We now take the  classical limit $\hbar \to 0$, keeping $\tilde W^{s}_m$ and $\tilde c  \sim {l \over  G}$ fixed, and replace the commutator by a Poisson bracket,
\be
\frac{1}{\hbar} \left[ \cdot \; , \cdot \right] \to \left\{ \cdot, \cdot \right\}
\ee
This yields what is known as the classical $W_N$ algebra in which all terms arising from normal orderings have disappeared, but which still contains nonlinear terms multiplied by powers of $1/\tilde c$.

In the second step, we neglect the terms multiplied by powers of $1/\tilde c$ as is appropriate in the weak gravity regime where we look at scales small compared to $\tilde c  \sim {l \over  G}$.  Doing this we obtain
 a linear algebra containing central terms proportional to $\tilde c$:
  \begin{align}\label{wtildecomm}
\left\{ \tilde L_m, \tilde L_n \right\} &= (m-n) \tilde  L_{m+n} +\left( {1\over 12} m(m^2-1) + 2m  h^{(-1)} \right)\tilde c \delta_{m,-n} \nonumber\\\
\left\{ \tilde L_m,  \tilde {W}^s_n \right\} &= \left((s-1)m-n\right) \tilde {W}^s_{m+n}+ s m w^{(-1)}_{s } \tilde c \d_{m,-n} \nonumber\\
\left\{ \tilde{W}^3_m, \tilde{W}^3_n \right\} &= 2(m-n)\tilde{W}^4_{m+n}+  (m-n)\left(\frac{1}{30}(2m^2+2n^2-mn-8) + {32\over 5 }h^{(-1)} \right)\tilde L_{m+n} \nonumber \\
  & \hspace{-60pt} + \left(\frac{1}{3 \cdot 5!}(m^2-1)(m^2-4) + 4 w^{(-1)}_{4 } + {32\over 5 }
 (h^{(-1)})^2 + {1 \over 15} (5 m^2 -8) h^{(-1)}  \right) m \tilde c \delta_{m,-n}+  \calo (1/\tilde c )\nonumber\\
\end{align}
Note that in this limit the structure constants  depend only on the leading part of the charges $w_s^{(-1)}$ as defined in (\ref{chargeslargec}).

In the perturbative sector we have $\L^-=0$ and hence the  terms involving to $w_s^{(-1)}$ are absent. We see from the above that in this case the central terms vanish for the wedge modes
$W^s_m$ with $|m|<s$ which form a closed  algebra in the limit \cite{Bowcock:1991zk} which we denote by $sl(N)_0$. Note that the wedge modes  are the modes which preserve the vacuum state $\ket{0}$.

We will now argue that also in the nonperturbative sector labeled by $\L^-$, there  are linear combinations of the generators which form, in the large $c$ limit, a closed $sl(N)$  algebra without central terms. We will denote this algebra by $sl(N)_{\L^-}$.
As in the perturbative sector, the algebra is constructed from the leading part of the  operators which preserve the primary $\ket{0,\L^-}$, in the sense that they produce a null state when acting on $\ket{0,\L^-}$. As we argued in section \ref{seclargec}, these operators are indeed linear combinations of the $W_N$ generators. The $W_N$  primaries $\ket{0,\L^-}$ should then behave as highest weight states of this  $sl(N)_{\L^-}$ algebra upon taking the large $c$ limit.  Indeed, we will argue that the state $\ket{\L^+,\L^-}$  corresponds simply to the highest weight of the $\L^+$ representation of the algebra $sl(N)_{\L^-}$.

 In what follows we will prove these statements  for the simplest  cases $N=2$ and $N=3$ and then discuss the
 expected generalization to arbitrary $N$. In the remainder of this section,  it will be convenient to label $sl(N)$ highest weights by $N-1$ Dynkin labels, which are related
 to the boxes in the Young diagram as $r_j - r_{j+1}$. We will denote the Dynkin labels of $\L^\pm$ by $d^\pm_j$, so that
\be
d^\pm_j = r^\pm_j - r^\pm_{j+1} \label{Dynk}.
\ee

\subsection{$N=2$ case}

Let's consider first the pure Virasoro case $N=2$. The charge $h^{(-1)}$ is given by
\be
h^{(-1)} = - {1 \over 24}  d^-_1  (d^-_1 +2) \label{O1hNis2}
\ee
where $d^-_1 $ denotes the Dynkin label of $\L^-$. We saw in (\ref{nullNis2}) that the representation built on $(0,\L^-)$ has a null vector at level $d^-_1  + 1$ given by
 \be
 \left( \tilde L_{-d^-_1  - 1} + \calo (\tfrac{1}{c}) \right) \ket{0, \L^-}.
 \ee
As discussed in section \ref{seclargec}, the leading part is linear in the generators while the suppressed terms contain composite operators. This leads us to introduce the following twisted wedge modes:
\bea
E & = & \frac{i}{d^-_1 +1} \tilde{L}_{d^-_1 +1} \nonu
F & = & \frac{i}{d^-_1 +1} \tilde{L}_{-d^-_1 -1} \nonu
H & = & -\frac{2}{d^-_1 +1} \tilde{L}_0
\eea
satisfying the $sl(2)$ algebra under the Poisson brackets defined in (\ref{wtildecomm}),
\bea
\left\{E, F \right\} & = & H \nonu
\left\{H, E \right\} & = & 2E \nonu
\left\{H, F \right\} & = & -2F\label{twistedsl2cft}
\eea
Now let's discuss the transformation properties of the primaries $|\L^+ , \L^- \rangle$ under $sl(2)_{\L^-}$. The state $|\L^+ , \L^- \rangle$  is clearly a highest weight state
under $sl(2)_{\L^-}$ and we can compute its weight:
\be
H \ket{\L^+ , \L^-} = d^+_1 \ket{\L^+ , \L^-}.
\ee
where $d^+_1$ is the Dynkin label of $\L^+$. From this and (\ref{hwsl3}) we see that $\ket{\L^+ , \L^-}$ is the highest weight of the $( d^+_1 + 1)$-dimensional irreducible representation $\L^+$ under $sl(2)_{\L^-}$
obtained by repeatedly acting with $F$.
An additional check that the representation is $(d^+_1 + 1)$-dimensional comes from the known large $c$ limit of the null vector in the $(\L^+ , \L^-)$ representation (\ref{nullNis2}), which is precisely
\be
\left(F^{d^+_1 + 1} + \calo (\tfrac{1}{c})\right) \ket{\L^+ , \L^-} \sim 0.
\ee

\subsection{$N=3$ case}

Now we turn to the case of the $W_3$ algebra, $N=3$, and consider representations $(\Lambda^+, \Lambda^-)$ with $\Lambda^-$ fixed.
The leading pieces of the spin 2 and spin charges $h$ and $w_3$ at large $c$, which determine the twisted wedge subalgebra, are
\bea
h^{(-1)} & = & -\frac{(d^-_1)^2 + (d^-_2)^2 + d^-_1 d^-_2 + 3d^-_1 + 3d^-_2}{72} \\
w_3^{(-1)} & = & i \frac{(d^-_1-d^-_2)(d^-_1+2d^-_2+3)(2d^-_1+d^-_2+3)}{324 \sqrt{10}}
\eea
We know from (\ref{nulllevels}) that the representation $(0, \Lambda^-)$ contains two independent null states at levels
\be
d^-_1+1 \quad\quad \text{and} \quad\quad d^-_2+1.
\ee
Furthermore, we expect from our discussion in section \ref{seclargec}  that these  null vectors are  linear in the level $d+1$ generators. Hence we make the following ansatz for
  the null vector at level $d+1$
\be
F \ket{0, \Lambda^-} \sim \left( \tilde{W}_{-d-1} + \alpha_{-d-1} \tilde{L}_{-d-1} \right) \ket{0, \Lambda^-},
\ee
The constant $\alpha_{-d-1}$ can be determined by requiring that the bracket of $F$ and its BPZ conjugate \footnote{Recall that the BPZ inner product is a bilinear form, so the BPZ conjugate flips the sign of mode indices without taking the complex conjugate of coefficients.}
\be
E \sim \tilde{W}_{d+1} + \alpha_{-d-1} \tilde{L}_{d+1}
\ee
has no terms proportional to the central charge (just as the central charge terms vanished for the wedge algebra generators for $\Lambda^-=0$). Having determined $\alpha_{-d-1}$, we find a set of generators satisfying the $sl(3)$  Lie algebra commutation relations in the Chevalley basis as given in Appendix \ref{appsl3}:
\begin{eqnarray}
E_1 &=& -i\frac{d^-_1+2d^-_2+3}{6\mathcal{N}\sqrt{(d^-_1+1)}} \tilde{L}_{d^-_1+1} + \frac{\sqrt{\frac{5}{2}}}{\mathcal{N}\sqrt{(d^-_1+1)}} \tilde{W}_{d^-_1+1} \nonu
E_2 &=& i\frac{2d^-_1+d^-_2+3}{6\mathcal{N}\sqrt{(d^-_2+1)}} \tilde{L}_{d^-_2+1} + \frac{\sqrt{\frac{5}{2}}}{\mathcal{N}\sqrt{(d^-_2+1)}} \tilde{W}_{d^-_2+1} \nonu
F_1 &=& -i\frac{d^-_1+2d^-_2+3}{6\mathcal{N}\sqrt{(d^-_1+1)}} \tilde{L}_{-d^-_1-1} + \frac{\sqrt{\frac{5}{2}}}{\mathcal{N}\sqrt{(d^-_1+1)}} \tilde{W}_{-d^-_1-1}\nonu
F_2 &=& i\frac{2d^-_1+d^-_2+3}{6\mathcal{N}\sqrt{(d^-_2+1)}} \tilde{L}_{-d^-_2-1} + \frac{\sqrt{\frac{5}{2}}}{\mathcal{N}\sqrt{(d^-_2+1)}} \tilde{W}_{-d^-_2-1} \nonu
H_1 &=& \frac{(d^-_1)^2-2(d^-_2)^2-2 d^-_1 d^-_2 -6d^-_2 -3}{3\mathcal{N}^2} \tilde{L}_0 -i\sqrt{\frac{5}{2}}\,\frac{d^-_1 + 2 d^-_2 +3}{\mathcal{N}^2} \tilde{W}_0 \nonu
H_2 &=& \frac{-2(d^-_1)^2 +(d^-_2)^2-2 d^-_1 d^-_2 -6d^-_1 -3}{3\mathcal{N}^2} \tilde{L}_0 +i\sqrt{\frac{5}{2}}\,\frac{2d^-_1 + d^-_2 +3}{\mathcal{N}^2} \tilde{W}_0\label{twistedsl3cft}
\end{eqnarray}
where we used a shorthand notation for the factor
\be
\mathcal{N} \equiv \sqrt{(d^-_1+1)(d^-_2+1)(d^-_1+d^-_2+2)}
\ee
Similarly to (\ref{twistedsl2cft}), one can check that these satisfy the correct $sl(3)$ commutation relations. To see how this $sl(3)_{\L^-}$ subalgebra acts on the $(\Lambda^+, \Lambda^-)$ representation of $W_N$ , we act with the Cartan elements $H_j$ on the highest weight state.
We find that the eigenvalues are simply the Dynkin labels of the highest weight $\L^+$:
\be
H_j \ket{\L^+,\L^-} = d^+_j \ket{\L^+, \L^-}.
\ee
This shows (see (\ref{hwsl3})) that acting on $\ket{\L^+, \L^-}$ with twisted wedge subalgebra generates the $\L^+$ representation of  $sl(3)_{\L^-}$ in the large $c$ limit. As in the $N=2$ case,  this would be consistent with large $c$ behavior of null states in the $(\L^+,\L^-)$ representation, i.e.  the null states in the large $c$ limit become
\be\label{largecnull}
F_j^{d^+_j+1} \ket{\L^+,\L^-} \sim 0.
\ee
We have checked that this property holds for some simple  $(\L^+,\L^-)$ representations in table \ref{nulltab} of Appendix \ref{appnull}.

\subsection{Summary and expected generalization}
Let's summarize the results of the last subsections and indicate the general  pattern which we expect to hold at all values of $N$.
From (\ref{nulllevels}) we know that the $(0, \Lambda^-)$ Verma module contains $N-1$ null vectors at levels $d^-_j+1$  which we write as
\be
\caln_j (0, \L^-) \ket{0, \L^-},\qquad  \ j = 1 \ldots N-1
\ee
We argued that the large $c$ limit
 \be
F_j \equiv \lim_{c \to \infty} \caln_j (0, \L^-)
\ee
is a linear combination of the $W_N$ generators. These linear operators play the role of simple roots, and
together with their   BPZ conjugates $E_j$ they generate the twisted wedge algebra $sl(N)_{\L^-}$  at large $c$  which annihilates $\ket{0,\L^-}$.

Now let's consider representations $(\Lambda^+, \Lambda^-)$ with $\Lambda^+ \neq 0$. From (\ref{nulllevels}) we know that they have $N-1$
independent  null states
occurring at levels $(d^+_j+1)(d^-_j+1)$. We write them as
 \be
\caln_j (\L^+, \L^-) \ket{\L^+, \L^-},\qquad  \ j = 1 \ldots N-1. \label{nulleqs}
\ee
In the large $c$ limit the operators $\caln_j (\L^+, \L^-)$ are no longer linear in the $W_N$ generators.
However, we found that they are simply  powers of the linear operators $F_j$, namely
\be
\lim_{c \to \infty} \mathcal{N}_j (\L^+, \L^-) = \left(  F_j  \right)^{d^+_j+1}.
\ee
The null state equations (\ref{nulleqs}) just tell us that $\ket{\L^+,\L^-}$ is the highest weight state of the  representation $\L^+$ of $sl(N)_{\L^-}$.

 As we will see in section \ref{secsymms}, these properties are rather manifest in the dual description of the theory, and hence the duality predicts them to hold for general $N$.
 We leave the CFT proof of these properties for general $N$  for future work.\footnote{As always, the results which we derived in this section in the holomorphic sector have an antiholomorphic
 counterpart, in particular there are additional antiholomorphic twisted wedge modes which generate a second symmetry algebra $\overline{sl(N)}_{\L^-}$ at large $c$.}

Having laid the CFT groundwork needed to establish the finer points of the semiclassical $W_N$ duality, we switch gears and study the relevant Vasiliev theory in the bulk.

\section{3d Vasiliev theory at $\l=-N$}

The proposed bulk dual to the 't Hooft limit of the $W_N$ minimal models is Vasiliev gravity with gauge algebra \hsl. If one thinks of the CFT semiclassical limit as an analytic continuation from the 't Hooft regime, this can be implemented by fixing $N$ finite and taking $k$ such that $c$ goes off to infinity. Triality of the quantum $W_{\infty}[\mu]$ algebra formally permits this continuation. Given that we have chosen the branch (\ref{klargec}), under this continuation one has $\l \rar -N$: therefore, one should study the Vasiliev theory based on Lie algebra hs[$-N]$ for integer $N$.

As we will discuss below, this is essentially a theory of flat $sl(N)$ connections -- containing non-propagating higher spin degrees of freedom with $s=2,3,\ldots,N$ -- with matter couplings consistent with higher spin gauge invariance. It is important to stress that the Vasiliev theory at $\l=-N$ and $sl(N)$ gravity are not equivalent, precisely because of the presence of other fields in the theory besides the flat connections.

The main result of this section is to argue that the field equations for linearized matter perturbations around classical backgrounds make sense at $\l=-N$. While there are extra constraints on scalar fields relative to the case of non-integer $\l$, their effect is to force the scalar into a non-unitary representation of the background isometry algebra. We propose to understand this holographically, via the corresponding non-unitarity on the boundary.

We first review the basic skeleton of the Vasiliev theory for general values of $\l$, at linearized order around flat background connections -- where much is understood -- and then specialize to $\l=-N$. Our discussion draws on \cite{Ammon:2011ua}.

\subsection{Basics of Vasiliev theory}

While the full 3d theory of higher spin gravity \cite{Prokushkin:1998bq} involves various mathematical objects required to write down fully nonlinear, higher spin symmetric interactions among higher spin fields and matter, it is essentially a theory of two flat connections coupled (quite non-minimally) to matter. These connections are valued in the Lie algebra \hsl, with generators $V^s_m$ with integer spin ($s\geq 2$) and mode ($|m|<s$) indices. In this language the identity is taken to be $V^1_0$, and the generators multiply via the associative `lone star product' \cite{Pope:1989sr},
\be
V^s_m \star V^t_n \equiv \half \sum_{u=1,2,...}^{s+t-|s-t|-1}g^{st}_u(m,n;\l)V^{s+t-u}_{m+n}~.
\ee
where $g^{st}_u(m,n;\l)$ are structure constants defined in e.g \cite{Gaberdiel:2011wb,Kraus:2011ds}, where more \hsl\ details are provided. The parameter $\l$ is an arbitrary number and parameterizes inequivalent algebras.

For generic $\l$, \hsl\ contains only an $sl(2)$ subalgebra, spanned by $\lbrace V^2_{\pm 1}, V^2_0\rbrace$ which obey
\be
[V^2_m,V^2_n] = (m-n)V^2_{m+n}
\ee
The rest of the \hsl\ generators can be built up from this $sl(2)$ via a universal enveloping algebra construction \cite{Gaberdiel:2011wb} in which $\l$ parameterizes the quadratic Casimir of the $sl(2)$ as $C_2 = {1\over 4}(\l^2-1)$; the objects \footnote{There is a second representation of \hsl\ in which one removes the factor of $(-1)^s$ from the generators.}
\be\label{env}
V^s_n = (-1)^{s-1-n} (4 q)^{s-2}
\frac{(n+s-1)!}{(2s-2)!} \,
 \Bigl[ \underbrace{V^2_{-1}, \dots [V^2_{-1}, [V^2_{-1}}_{\hbox{\footnotesize{$s-1-n$ terms}}}, (V^2_1)^{s-1}]]\Bigr] \ .
\ee
generate \hsl $~\oplus ~\CC$, where $\CC$ is the identity $V^1_0$ (and $q$ is a free normalization constant). It is easy to see that the structure constants will be polynomials in $\l^2$. Aside from integer $\l$ which we discuss at length below, the value $\l=1/2$ is special in that generators can be written as monomials in oscillators which multiply via the Moyal product.

The flat \hsl\ connections, then, obey the following equations:
\bea\label{flatc}
F &=&d A + A \wedge \star A =0\nonu
\bar F &=& d \bar A + \bar A \wedge  \star\bar A =0.
\eea
In a Lorentzian theory, these are independent and the gauge algebra is \hsl$\oplus$\hsl. One can consider the Euclidean theory, in which case it is natural to consider one complex connection valued in a complexified \hsl, which is what we do henceforth.

The matter is encoded in the `master field' $C$, a spacetime 0-form valued in \hsl $~\oplus ~\CC$, which obeys the elegant equation
\be\label{matter}
dC + A\star C - C\star \Ab=0
\ee
Satisfaction of this equation, together with (\ref{flatc}), constitutes only part of a (perhaps linearized) solution to the Vasiliev theory which contains a third master field, but (\ref{matter}) is all one needs to describe the coupling of matter to the higher spin fields.

The equations (\ref{flatc}) and (\ref{matter}) are invariant under the finite gauge transformations
\bea\label{gauget}
A &\rar& g^{-1}\star (A+d)\star g\\
\Ab &\rar& \overline{g}\star(\Ab+d)\star\overline{g}^{-1}\\
C &\rar& g^{-1}\star C \star \overline{g}^{-1}
\eea
To extract the physical content of these equations, we expand in \hsl$~\oplus~  \CC$,
\bea
C &=&\sum_{s=1}^\infty \sum_{|m|<s} C^s_m(x^\mu) V^s_m~, \nonu
A &=&\sum_{s=2}^\infty \sum_{|m|<s} A^s_m(x^\mu) V^s_m~, \nonu
\Ab &=&\sum_{s=2}^\infty \sum_{|m|<s} \Ab^s_m(x^\mu) V^s_m~,
\eea
and perform lone star products. The physical scalar field is the identity component of $C$:
\be
\Phi = C^1_0(x^{\mu})
\ee
The higher spinorial components of $C$ can be solved in terms of spacetime derivatives of $\Phi$.

Using this method, it is easily shown that one recovers the Klein-Gordon equation for a scalar in AdS, with mass $m^2 = -1+\l^2$. (We set the spacetime length scale to $\ell=1$ from now on.) In \cite{Ammon:2011ua}, this method was utilized to extract generalized wave equatons for scalar fields in higher spin backgrounds. In fact, one can obtain their solutions directly -- bypassing the need to find the relevant equation in the first place -- by making use of the gauge transformations (\ref{gauget}). This approach was also recently used in \cite{Kraus:2012uf} to compute bulk-boundary scalar propagators in higher spin spacetimes, including the higher spin black hole of the \hsl\ theory \cite{Kraus:2011ds}.

For generic $\l$, there is an infinite number of components $\lbrace C^s_m\rbrace$, corresponding to the infinite dimensionality of \hsl. The resulting wave equations for $\Phi$ are free field equations for this reason. The situation changes at integer $\l=\pm N$, where an ideal forms in \hsl\ and one has the option to truncate the algebra to those generators with $s\leq N$. Our goal is to understand the extra constraints that come from having only a finite tower of components of $C$.

\subsection{The special case $\l = \pm N$}

When  $\l = \pm N$ for integer $N$, the quadratic form on \hsl\ becomes degenerate and the  generators with vanishing norm form an ideal. After quotienting by the ideal , the remaining generators form an $sl(N)$ algebra, and, if we work in the $N$-dimensional representation, the lone star product becomes the ordinary matrix product. We will specialize to the Euclidean theory, in which case the complexified \hsl\ becomes, upon modding out by its ideal, $sl(N,\CC)$. All of our results of the previous subsection hold upon making this matrix substitution: in particular, the $sl(N,\CC)$ is generated by (\ref{env}), and the field equations become ordinary matrix equations. The identity $V^1_0$ becomes
\be
V^1_0 = {1 \over 4 q} {\bf 1}_{N\times N}.
\ee
We now expand the master field $C$ in generators\footnote{We note in passing that one need not mod out the ideal. See e.g. \cite{Gaberdiel:2012ku} for a recent discussion of one implication of this. In that case, one can still meaningfully expand $C$ in an infinite number of components, and there is some indication that the components with $s>N$ may hold some physical content. This has yet to be definitively understood. See \cite{Prokushkin:1998bq} for original discussions of this issue.} only up to spin $N$:
\be
C  =  {4 q\over N} \sum_{s = 1}^{N}\sum_{|m|<s} C^s_m(x^{\mu}) V^s_m.
\ee
In other words, the master field $C$ takes values in $sl(N) \oplus \CC$: it is simply an arbitrary complex $N\times N$ matrix. It is straightforward to identify the physical scalar field as the ordinary trace of $C$:
\be
\Phi = \tr (C)
\ee

We choose the Hermitian conjugation convention
\be
\Ab = -A^{\dagger}
\ee
The $sl(N)$ generators obey the conjugation relation
\be
(V^s_m)^{\dagger} = (-1)^m V^s_{-m}
\ee
To return to the metric-like formulation, one can define the vielbein and spin connection
\bea
e &=& {1 \over 2} (A +A^{\dagger} )~, ~~ \o = {1 \over 2i } (A -A^{\dagger} ).\label{vielbdef}
\eea
and he metric can be written as
\bea ds^2 &=& {12 \over N(N^2-1)} \tr e^2 .\label{metr}
\eea

Summarizing, in terms of $A, A^{\dagger}$, the field equations we will study henceforth are as follows:
\bea
&&dA + A\wedge A = 0\\
&&dA^{\dag} - A^{\dag}\wedge A^{\dag} = 0\\\label{slneqn}
&&dC + AC + CA^{\dagger}=0
\eea
with a gauge invariance (\ref{gauget}), where $\Ab = -A^{\dag}$ and $\overline{g} = g^{\dag}$.

Following previous work on the subject, we will choose a gauge in which the fields are independent of the radial coordinate. Labeling the radial coordinate $r$ and the boundary lightcone coordinates ($z,\zb)$ with $z\equiv \phi+i t_E$, we take
\bea\label{fgauge}
A = b^{-1}(a+d)b~, ~~b=e^{rV^2_0}
\eea
which acts as $b^{-1}V^s_mb = e^{mr}V^s_m$ on generators. For a translation invariant solution, $a$ is a constant 1-form, hence the flatness condition is merely $[a_z,a_{\zb}]=0$. Flat connections (\ref{fgauge}) give rise, upon decomposing (\ref{slneqn}) along spacetime, to matter equations
\bea\label{mattereqs}
\partial_r C &=& -\lbrace V^2_0,C\rbrace\\
\partial_z C &=& -(b^{-1}a_z b )C - C(ba_{\zb}^{\dag}b^{-1}) \\
\partial_{\zb}C &=& -(b^{-1}a_{\zb}b )C - C(ba_{z}^{\dag}b^{-1})
\eea

We proceed to study these equations in AdS spacetimes, both global and Poincar\'e. They will be shown to describe a Klein-Gordon scalar field with the expected mass $m^2 = -1+N^2$, subject to additional constraints that force it into a non-unitary highest weight representation of the AdS isometry algebra. Classification of these representations was recalled in \cite{Balasubramanian:1998sn} where scalar solutions in the bulk were shown to furnish such representations, and we will reproduce the results of \cite{Balasubramanian:1998sn} exactly. (See also \cite{Maldacena:1998bw}.)

Motivated by holographic correspondence with the semiclassical $W_N$ CFT, we will focus on the scalar field in standard quantization, with chiral conformal weights
\be
h=\overline{h} = {1+\l\over 2}~.
\ee
As argued earlier, we want to take the branch $\l=-N$, in which case these conformal weights become negative in the region of interest, $N\geq 2$.

\subsection{Scalars in global AdS}

The global AdS connection can be written as
\be\label{gadsc}
a = \half(V^2_1 + V^2_{-1}) dz\equiv a_{AdS} dz
\ee
In this background, the matter equations reduce to
\bea
 \pa_r C &=& - \lbrace V^2_0 ,C \rbrace\label{rhoeq}\\
\pa_z C &=& - (b^{-1} a_{AdS} b)  C \label{zeq}\\
 \pa_{\bar z} C &=& C(b a_{AdS}b^{-1}) \label{zbeq}
 \eea

We will soon establish a nice way to solve these equations for all $N$, but first we treat a simpler example by brute force. Specializing to $\l=-2$, this is $sl(2,\CC)$ gravity coupled  to matter in a specific way. Expanding $C$ in components $\lbrace \Phi, C^2_{\pm 1}, C^2_0\rbrace$, one finds that $\Phi$ obeys the Klein-Gordon equation along with constraints which give a four-parameter family of solutions:
\bea\label{gads}
\Phi &=& \a_1 e^{{i\over 2}(z+\zb)}\sinh(r) + \a_2 e^{-{i\over 2}(z+\zb)}\sinh(r)\\ &+&\a_3 e^{{i\over 2}(z-\zb)}\cosh(r) +\a_4 e^{-{i\over 2}(z-\zb)}\cosh(r)
\eea
where the $\a_i$ are constants. The $C^2_m$ components are linear in derivatives of $\Phi$, e.g. $C^2_0 = -2\pa_r \Phi$.

These four solutions are, in fact, recognizable as those of the non-unitary $(2,2)$ representaiton of the $sl(2) \times \overline{sl(2)}$  isometry\footnote{We work with the Euclidean theory, hence global AdS is invariant under the $sl(N,\CC)$ global symmetry. However it will be useful to think in terms of left and right symmetry algebras; one should keep in mind that they are related by conjugation and form a single $sl(N,\CC)$.} of global AdS. It was established in \cite{Balasubramanian:1998sn} that solutions of the Klein-Gordon equation in global AdS furnish highest weight representations of this symmetry, filled out by acting on the highest weight state with differential operator representations of the symmetry generators. That this is true follows from the fundamental role of symmetry in AdS/CFT. Having taken $\l=-2$, we have $h=\overline{h} = -\half$; from \cite{Balasubramanian:1998sn} (see section 4.1 therein), this puts us in the (2+2)-dimensional, irreducible representation labeled ${\cal{D}}(-\half)\times {\cal{D}}(-\half)$.

Let us elaborate on this after solving the scalar equations for arbitrary $N$. We use a method which naturally organizes the representation theoretic content of the solutions. The global AdS connection (\ref{gadsc}) is diagonalizeable to a matrix $D$  given by
 \be
D= - i {\rm diag}( \r_i )  = -i V^2_0 , \qquad i = 1, \ldots , N.
 \ee
A convenient choice for the similarity transformation which diagonalizes $a_{AdS}$ is
\be
M = e^{-{i \pi \over 4} (V^2_1 - V^2_{-1})}.
\ee
using which one finds $a_{AdS} = M^{-1} D M$. Note that $M^{\dag}=M^{-1}$. To solve the matter equations, we first construct solutions in this diagonal gauge and then transform back to the physical gauge. In the diagonal gauge, call the master field  and the gauge field $\tilde C$ and $\tilde A$, respectively. Then these two gauges are related by
\bea
A &=& (Mb)^{-1}( \tilde A+d)(Mb)\\
C &=& (Mb)^{-1} \tilde C M b^{-1}
\eea
The scalar equations in the diagonal gauge become simply
\bea
\pa_z \tilde C &=&  i V^2_0 \tilde C\\
\pa_{\bar z} \tilde C &=&  -i \tilde C  V^2_0. \label{eqsdiaggauge}
\eea
 There are  $N^2$ independent solutions given by
\be
\tilde C_{ij} =  e^{i (\r_i z -  \r_j \bar z)} e_{ij}\qquad i,j = 1,\ldots , N. \label{soldiaggauge}
\ee
where $e_{ij}$ are the standard basis of $N\times N$ matrices: $(e_{ij})_{kl} = \d_{ik} \d_{jl}$. Transforming back to the physical gauge and taking the trace we find  $N^2$ linearly independent solutions for $\Phi$:
\bea
\Phi_{ij} &=& e^{i (\r_i z -  \r_j \bar z)} P_{ij} (r)\qquad i,j = 1,\ldots , N\\
P_{ij} (r) &=& \left( M e^{-2 r V^2_0} M^{-1} \right)_{ji}
\eea
We can work out the $r$-dependent part $P_{ij} (r)$ of the solution explicitly:
\bea
P_{ij} (\r)  &=&  \left( M e^{-2 r V^2_0} M^{-1} \right)_{ji} = \left(  e^{- i r (V^2_1 + V^2_{-1})} \right)_{ij}= \left(  e^{ 2 r T_2} \right)_{ij}
\eea
In the last step, we have used that the combination $- i (V^2_1 + V^2_{-1})/2$ can be seen as the $sl(2)$ generator $T_2$ in the $N$-dimensional representation.
Its exponential can be
expressed in terms of Jacobi polynomials which gives (up to overall normalization),
\bea
P_{ij} (r)  &~& (\sinh r)^l (\cosh r)^\o P^{(l,\o)}_{\half( N-1 -l -\o )} (\cosh 2 r )\\
&\sim& (\sinh r)^l (\cosh r)^{N-1-l} \ _2F_1 \left( \half(l +\o + 1 -N),\half(l -\o + 1 -N),l+1, \tanh^2 r\right)\nonumber
\eea
where $P^{(\a,\b)}_n$ denote the Jacobi polynomials and $\o, l$ denote the frequency and angular momentum:
\bea
\o &=& |N+1- (i+j)|\\
l &=& |i-j|.
\eea
These are the solutions of \cite{Balasubramanian:1998sn}. One can quickly check that this reproduces our results for $N=2$.

We argued in the previous section that the global AdS background is invariant under (two copies of) an $sl(N)_0$ algebra of transformations, and hence the fluctuations of the scalar field should be organized into its irreducible representations. Let us start with the diagonally embedded
$sl(2)_0 \in sl(N)_0$ subalgebra. As we show in Appendix B.2 -- and as we will discuss in more detail in section 5, when we put scalars atop surpluses -- global AdS admits symmetries which, acting on scalar solutions $\Phi_{ij}$, reduces to the standard action of the Killing vectors of AdS$_3$ on scalars.
The differential operators
\bea
l_0 &=& i \pa_z \\
l_{\pm 1} &=& i e^{\pm i z}\left( \coth 2 r \pa_z + {1 \over \sinh 2r} \pa_{\bar z} \mp {i \over 2} \pa_r\right)\label{KVs}
\eea
together with the $\bar l_m$ obtained by complex conjugation, furnish a canonically normalized $sl(2) \times \overline{sl(2)}$ algebra.

The solution $\Phi_{11}$ has the lowest energy and is a highest weight state:
\be\label{hws}
\Phi_{11} = e^{ih(z-\zb)}{1\over (\cosh r)^{2h}}
\ee
where $h=\bar h = {1-N\over 2}$ are the left and right excitation energies \cite{Maldacena:1998bw, Balasubramanian:1998sn}. It is manifestly primary, $l_1\Phi_{11}= \bar l_1 \Phi_{11}=0$. This has the right properties to be identified with the primary $|f,0\rangle$ in the dual CFT. The other scalar solutions are related to the primary solution by repeatedly acting with the creation operators as $\Phi_{ij} = (l_{-1})^{i-1} (\bar l_{-1})^{j-1} \Phi_{11}$, and have energies $l_0 = h+i-1, \bar l_0 = h+j-1$: these are dual to the descendants of $|f,0\rangle$ filling out the non-unitary representation ${\cal{D}}(h)\times {\cal{D}}(h)$. The property $(l_{-1})^{N}  \Phi_{11}= (\bar l_{-1})^{N}  \Phi_{11}=0$ corresponds to the
large $c$ behavior (\ref{largecnull}) of the null descendants in the dual theory, i.e. the representation is finite-dimensional. One can confirm that the $sl(2)_0$ Casimir takes the value
\be
\left[ (l_0)^2 - \half ( l_1 l_{-1} + l_{-1} l_1 )\right] \Phi_{ij} = {1\over 4} (N^2-1) \Phi_{ij}\label{Cas}
\ee
which implies that the scalars all have the expected mass, $m^2=-1+N^2$.

The entire preceding discussion -- including the solution (\ref{hws}) and its descendants -- matches the results of \cite{Balasubramanian:1998sn}. From a holographic viewpoint, the non-unitarity of the scalar representation reflects the non-unitarity in the proposed CFT dual to this theory. For $N>2$ we can go further, by understanding how the scalar solutions transform under the full $sl(N)_0$ symmetry of the background. Since the $sl(2)_0$ subalgebra considered above is diagonally embedded and the solutions form an $N$-dimensional representation of $sl(2)_0$, the action of the $sl(N)_0$  generators can be constructed as elements of the enveloping algebra
    as in (\ref{env}), with the spin-$s$  generators acting as order $s-1$ differential operators. The space of solutions
    transforms as a fundamental representation under this $sl(N)_0$ action, consistent with the properties of the $(f,0)$ representation in the dual CFT.
\vs
Based on this evidence we conclude that the space of scalar single particle states in global AdS corresponds to the CFT primary $|f, 0\rangle$ and a subset of descendants which are obtained by acting with wedge modes. As we should expect from other studies of linearized scalar wave equations in 3d Vasiliev theory \cite{Chang:2011mz, Ammon:2011ua, Kraus:2012uf}, this is determined by the \hsl\ $\rar sl(N,\CC)$ symmetry respected by the vacuum.

In Appendix B, we provide explicit details on the constraints which force the scalar into the non-unitary representation, as well as a demonstration that this programme works in Poincar\'e AdS equally well.

\section{Matching bulk solutions to CFT representations}
Having obtained an understanding of the boundary symmetries and bulk equations in AdS, we now make the central arguments for matching CFT representations in the large $c$ limit with conical surpluses and scalar excitations thereof. We begin with a short review of the surplus solutions in the Vasiliev theory \cite{Castro:2011iw}.

\subsection{Review: smooth solutions on the solid cylinder}\label{secsurplus}
To begin, we must review the standard boundary conditions on the gauge connections which define asymptotically  AdS solutions. As shown in \cite{Campoleoni:2010zq}, these boundary conditions lead to a classical $W_N$ algebra of asymptotic symmetries, which matches the symmetries of the proposed CFT dual at large $c$.

As explained in \cite{Campoleoni:2010zq}, after imposing AdS  boundary conditions and fixing the so-called ``highest weight'' gauge, the allowed flat connections are of the form
\bea
A &=& b^{-1} a(z) b dz + b^{-1} db \qquad b = e^{r V^2_0} \label{rhodepgauge} \\
a(z) &=& \half V^2_1 + {12 \p \over c} \sum_{s = 2}^N {2^{s-1}\over N_s} W_s (z) V^{s}_{-(s-1)}\label{hwgauge}
\eea
where the $N_s$ are suitably chosen normalization constants\footnote{Our conventions  follow \cite{Gaberdiel:2011wb}, where $N_s = {6\over N(N^2-1)} \tr V^s_{-s+1} V^s_{s-1}$.The explicit expression is $N_s ={3 \cdot 4^{s-3}\sqrt{\pi}q^{2s-4}\Gamma(s)\over (N^2-1)
\Gamma(s+\half)} (1-N)_{s-1} (1+N)_{s-1} $
where $(a)_n = \Gamma(a+n)/\Gamma(a)$ is the ascending Pochhammer symbol. In particular, we have $N_2 = -1$.}.

We can expand the holomorphic functions $W_s(z)$ in Fourier modes on the cylinder:
 \be
 W_s (z)= {1\over 2 \p} \sum \left( W^s_n - {c\over 24} \d_{s,2} \d_{n,0}\right) e^{- i n z}. \label{modeexp}
 \ee
 The modes $W^s_n$ parameterize the classical phase space of the massless higher spin theory. Computing their Poisson brackets, one finds that the $W^s_n$  generate the classical limit of the $W_N$  algebra  (\ref{wcomm}) \cite{Campoleoni:2010zq}. This computation will be reviewed in section \ref{secsymms} below. The modes are
normalized as in (\ref{wcomm}) if we take the normalization constant $q$ in (\ref{env}) to be
\be
q = {1\over \sqrt{8 (N^2-4)}}.
\ee
This is the value we adopt from now on.

The conical surplus solutions constructed in \cite{Castro:2011iw} are smooth classical geometries which obey the boundary conditions (\ref{hwgauge}), asymptotic to global AdS. These will act as the
saddle points in the semiclassical expansion of the bulk path integral. The global AdS solution has the topology of a solid cylinder, with the Euclidean time
running along the length of the cylinder; we fix this to be the topology
of the 3-manifold on which the theory is defined, and our CFT lives on the boundary of the solid cylinder. Choosing again our complex coordinate to be $z = \f + i t_E$, with $\f$ an angular coordinate with periodicity $\f \sim \f + 2 \p$, the $\f$-circle is contractible in the bulk.

We restrict attention to solutions which preserve time translation and rotational invariance, which means that we look at connections of the form (\ref{rhodepgauge}) with $a(z)$ a constant Lie algebra element; in anticipation of the match to CFT representations, we call the $z$-component of these constant connections $a_{\L^-}$. The requirement that the Chern-Simons gauge field is smooth imposes that its holonomy  $H$ along the contractible $\f$-cycle is a trivial element of the gauge group.
In order to clarify what we mean by  `trivial', we have to be a bit more specific  about the global structure of the gauge group. The group  $SL(N,\CC)$
has a $\ZZ_N$ center generated by elements of the form $e^{ 2 \p i m/N} {\bf 1}$ with $m$ integer. As we can see from (\ref{gauget}), the central elements act trivially on the gauge fields $A,\bar A$ and as well as on the matter field $C$. Therefore the actual gauge group is  $SL(N,\CC)/\ZZ_N$.
The condition of trivial holonomy then imposes
\be
H \sim e^{2 \p a_{\L^-}} = e^{  2 \p i m/N} {\bf 1}\label{holcond}
\ee
This means that the
eigenvalues $\l_i$ of  $a_{\L^-}$ are imaginary, $\l_i = - i n^-_i$ with
\be
n^-_i = m_i - {m \over N}; \qquad m_i \in \ZZ; \qquad i = 1, \ldots , N.\label{trivhol}
\ee
The requirement that the sum of the $n^-_i$ vanishes imposes  that  $m = \sum_j m_j$.

We have yet to impose the boundary conditions, which require that $a_{\L^-}$ is of the form (\ref{hwgauge}). As shown in \cite{Castro:2011iw}, matrices
of this form necessarily have eigenvalues which are all distinct. Without loss of generality, we can assume the $n^-_i$, and hence also the $m_i$, to form a strictly ordered
set: $m_1 > m_2 > \ldots > m_N$. We will now show that the information contained in the $m_i$ encodes precisely an $sl(N)$ Young diagram.

Firstly, we see from (\ref{trivhol})  that the $m_i$ are only determined up to an overall shift $m_i \to m_i + p, \ p \in \ZZ$. We use this freedom to fix
$m_N =0$. It is then easily seen that the numbers $r^-_i$ defined by
\be
r^-_i = m_i - (N-i)
\ee
satisfy $r^-_1 \geq r^-_2 \geq \ldots \geq r^-_N =0$, in other words the $r^-_i$ are in one-to-one correspondence with Young diagrams of $sl(N)$. Substituting into (\ref{trivhol})
gives
\be
n^-_i = r^-_i - {B^- \over N} + {N+1\over 2} - i = \L^-_i + \r_i
\ee
where $B^-= \sum_i r^-_i$ is the number of boxes in the Young diagram specified by the $r^-_i$, $\L^-_i$ is the highest weight vector determined
by the Young diagram and $\r$ is the Weyl vector (see (\ref{defLambda}), (\ref{defWeyl})).

For example, for global AdS, $a$ is given by $a_{AdS} = \half (V^2_1 + V^2_{-1})$ and  the eigenvalues are
\be n^-_{i, AdS} = {N+1 \over 2} - i = \r_i \label{AdSns}\ee
corresponding to the trivial $sl(N)$ representation with $r^-_i =0$.

As another example, consider the smooth solutions of the $N=2$ theory. These are labeled by a single natural number $r^-_1$ and the eigenvalues are
$n^-_1 = (r^-_1+1 )/2, n^-_2 = -(r^-_1+1 )/2$, corresponding to
taking
\be a_{\L^-} = \half (V^2_1 + (r^-_1+1 )^2 V^2_{-1}).\label{Nis2surplus}\ee
Computing the metric using (\ref{metr}) one finds
\be
ds^2 =  dr^2 + (r^-_1+1 )^2 \cosh^2 rdt_E^2 +  (r^-_1+1 )^2 \sinh^2 rd\f^2.
\ee
When $r^-_1$ vanishes, the metric is smooth and represents global AdS.  For nonzero $r^-_1$, the metric has a conical singularity at $r=0$ with opening angle
 $2 \p (r^-_1+1 )$, hence the name `conical surplus'.\footnote{Of the general smooth solutions under consideration, only a subset (namely the ones where the nonvanishing  $n^-_i$ come in opposite pairs) correspond to
conical surplus metrics when written in a specific gauge \cite{Castro:2011iw}. Nevertheless we will be a bit imprecise in what follows and refer to all the smooth solutions as `surpluses'.}

The higher spin charges these surpluses are the coefficients in the expansion (\ref{hwgauge}).
By demanding smooth holonomy, these are fixed upon requiring that the matrix $a_{\L^-}$ in  the highest weight gauge (\ref{hwgauge}) has the same trace invariants as the diagonal matrix with eigenvalues $- i n^-_j$. So far the general expression for the charges is not  known but they can be easily computed recursively to arbitrary order. The first few charges are \cite{Castro:2011iw}
 \bea\label{surplcharge}
 W^2_0  &=&- {c \over N(N^2-1)} C_2(n^-)~,\nonumber\\
 W^3_0 &=& { \sqrt{2} i  c\over N(N^2-1)\sqrt{N^2-4}} C_3 (n^-) ~,\nonumber\\
 W^4_0 &=&
  { 2 c\over N(N^2-1)(N^2-4)} \left( C_4 (n^-) - {C_4(\r) \over C_2(\r)^2 } C_2(n^-)^2\right)~.\label{charges}
 \eea
 where, as before, $C_s (n) = {1 \over s} \sum_i (n_i)^s$.
Comparing to
 (\ref{O1weight}), (\ref{O1spin3}) we see that these charges are
 consistent (upon choosing the branch  $\sqrt{4-N^2} = i\sqrt{N^2-4}$) with the identification of the  surpluses with CFT primaries of the form $(\L^+,\L^-)$ for some representation $\L^+$. From the classical charges alone we cannot determine $\L^+$, since we know from (\ref{O1weight}), (\ref{O1spin3}) that for fixed $\L^-$ all primaries $(\L^+ ,\L^-)$ have degenerate charges in the large $c$ limit. However, we will argue in the next section from an analysis of the null vectors that the correct identification of the surpluses is with the primaries $(0, \L^- )$.

We conclude this section with an important remark. In Chern-Simons theories the gauge connection can always locally  be   written in a pure gauge form
$A = g^{-1} d g$.
For the surplus solutions the gauge parameter is
\be
g_{\L^-} =  e^{ a_{\L^-} z}.\label{puregauge}
\ee
The condition of trivial holonomy (\ref{holcond}) actually ensures that the gauge parameter respects the $\f$ periodicity and is hence globally well-defined. The  fact
that surplus connections are pure gauge in the Chern-Simons sense will be useful later.
We should keep in mind however that the Chern-Simons gauge transformations  which relate the various surpluses are  large  gauge transformations which   act nontrivially on the boundary
and relate inequivalent classical solutions with different charges.

\subsection{Boundary higher spin particles and symmetries}\label{secsymms}
Since the higher spin fields are described by flat gauge connections,  all on-shell fluctuations of the higher spin fields are pure gauge. However, in the presence of a boundary, only the gauge parameters which fall off fast enough near the boundary should be regarded as true gauge transformations.  Gauge parameters which act on the boundary but respect the boundary conditions (in our case, (\ref{hwgauge}))  should
be regarded as  asymptotic symmetry transformations which relate inequivalent classical configurations. As already mentioned before, the corresponding conserved charges generate a classical $W_N$ algebra under Poisson brackets, which we now review in some more detail.

It will be convenient to work in the `$r$-independent gauge' where the gauge connection is simply given by $A = a(z)dz$. We can transform back to the $r$-dependent gauge
  (\ref{rhodepgauge}) by making a gauge transformation with parameter $b = e^{r V^2_0}$.
  The parameters $\z$ generating asymptotic symmetries are determined by requiring that, for  $a$ in the highest weight gauge (\ref{hwgauge}),
the gauge-transformed connection $\d_\z a = \z' + [a, \z]$ is still of the form (\ref{hwgauge}).
As shown in \cite{Campoleoni:2010zq}, for every spin $s=2,\ldots,N$, there is  a  gauge parameter $\z^s$ that depends
on an arbitrary  holomorphic function $\h_s (z)$ and is of the form
\be
\z_s ( z) ={\h_s (z) \over 2^{s-1}}  V^s_{s-1} +\ldots
\ee
where $+ \ldots$  stands for terms involving non-lowest weight generators of $sl(N)$ with coefficients determined by $\h_s$ and its derivatives.
A symmetry generator $\z$ is associated to a conserved charge $Q[\z]$ which generates the corresponding symmetry under Poisson brackets:
\be
Q[\z^s] = \int_0^{2 \p} dz   \h_s (z) W_s (z)
\ee
such  that
\be
\d_{\z_s} Q[\tilde \z_t  ] = \{ Q[\tilde \z_t ] , Q[\z_s  ]\}_{PB}. \label{convertPB}
\ee
We will denote by $\z^s_n$ the gauge parameters such that the associated charges are $Q[ \z^s_n ]  = W^s_n - {c\over 24} \d_{s,2} \d_{n,0}$. From (\ref{modeexp}) we see that these are obtained by taking $\h_s = e^{inz}$.

For example, the spin 2 and spin 3 gauge parameters are given by
\bea
\z_2 &=& \h_2 a - \h_2' V^2_0 + \h_2''  V^2_{-1}\nonu
\z_3 &=& 4 q \left( a^2 - {{\rm tr} a^2\over N} V^1_0\right)\eta_3 -{\h_3'\over 2} V^3_1 + {\h_3'' \over 2} V^3_0  + \left( { 8 \p \over c} ( 2 \h_3 W_2'+ 5 W_2 \h_3') - {\h_3^{'''} \over 3}\right) V^3_{-1}\nonu
&& - \left( { 4 \p \over c} ( 7 W_2'\h_3'+ 2 \h_3 W_2'' + 8 W_2 \h_3'')- {\pa^4\h_3 \over 6}\right) V^3_{-2} + \ldots \label{gaugepars}
  \eea
In the last line we have omitted terms involving $sl(N)$ generators of spin 4 and higher, i.e. the expression is exact only for $N=3$.

The stress tensor $W_2$ and higher spin currents $W_s$ transform under the the spin 2 and spin 3 transformations as
\bea
\d_{\z_2} W_2 &=& 2 \h_2' W_2 +  \h_2 W_2' - {c \over 24 \p} \h_2'''\nonu
\d_{\z_2} W_s &=& s \h_2' W_s +  \h_s W_s' \qquad s>2 \label{spin2transfo}\\
\delta_{\z_3} W_2 &=& 3 W_3 \h_3' + 2 W_3' \h_3\nonu
\delta_{\z_3} W_3 &=&  4 W_4 \eta_3' + 2W_4'\eta_3-{1\over 30}(15W_2'\eta_3''+9\eta_3'W_2'' + 10 W_2 \eta_3''' + 2 W_2'''\eta_3)  \label{com33} \\
& & \quad + {64\pi\over 5c} W_2(W_2\eta_3)' + {c\over 720 \pi} \pa^5\eta\notag.
\eea
Converting to Poisson brackets using (\ref{convertPB}) and expanding in modes using (\ref{modeexp}) these transformations reduce to the classical limit of the commutation relations (\ref{wcomm}).

So far we have, for a given background $a$, constructed a space of linearized fluctuations of the form $\d_{\z^s_n} a$. To select which of these fluctuations represent single particle states, we should identify a suitable subspace
of `positive frequency' solutions\footnote{In standard unitary quantum field theory this is achieved by selecting normalizeable, positive norm modes with respect to a canonical inner  product. Even though a canonical inner product exists for higher spin fluctuations, see \cite{Castro:2011ui}, we cannot apply this method here since we are considering a non-unitary theory.} or, in other words, choose a vacuum. The natural prescription from the point of view of the underlying $W_N$ symmetry is to associate single particle states with the fluctuations with negative mode number, $\d_{\z^s_{-n}}a$ with $n>0 $. For $s=2$ these are usually called boundary gravitons, and for general spin we will refer to them as boundary higher spin particles.

The gauge field fluctuations $\d_{\z^s_{-n}}a$  are generically linearly independent and hence represent distinct single particle states. This is however not necessarily true
  when the background has a symmetry. Indeed, a symmetry  is by definition a gauge transformation which leaves the solution invariant. Expanding the corresponding gauge parameter  in modes $\z^s_{n}$, we find a linear combination of fluctuations  which vanishes. If the mode numbers in this  combination  are negative not all  $\d_{\z^s_{-n}}a$ represent distinct boundary particles. It is intuitively clear that such classical symmetries in the bulk are closely related to the appearance of null states in the Verma module in the dual CFT.
For example, in the global AdS background it is easy to see  that $\z^2_0, \z^2_{\pm 1}$ generate symmetries i.e.
the right-hand side of (\ref{com33}) vanishes. This simply reflects the $sl(2)$ symmetry of the AdS background. Hence the fluctuation $\d_{\z^2_{-1}} a_{AdS}$
does not correspond to a boundary graviton. The corresponding CFT statement is that $L_{-1}|0\rangle$ is a null state at large $c$.

As a more nontrivial example, consider the surplus  specified by $\L^- = f$ at general $N$. From (\ref{charges}) we find its classical charges
\bea
W^2_0 &=& - {c \over 2 N^2} - {c\over 24}\nonu
W^3_0 &=& {i \sqrt{N^2-4}\over 3 \sqrt{2} N^3} c\nonu
W^4_0 &=& {N^2-9\over 10 N^4} c.
\eea
Using these in the transformation rules (\ref{com33}), we can look for candidate symmetries of the background $\z_\caln$ which are linear combinations of the spin-2 and spin-3
generators and satisfy $\d_{\z_\caln} W^2 = \d_{\z_\caln} W^3=0$. One easily checks that such combinations exist at level one and two and are given by:
\bea
\z_{\caln,1} &=& \z^3_{-1} + {i \sqrt{N^2-4} \over \sqrt{2} N} \z^2_{-1}\nonu
\z_{\caln,2} &=& \z^3_{-2} - {2 \sqrt{2} i \over N \sqrt{N^2-4} } \z^2_{-2}.
\eea
These combinations correspond precisely to the level one and two null states in the $(0,f)$ representation of the CFT at large $c$, see (\ref{fnullvects}), if we again choose the branch
$\sqrt{4-N^2} = i\sqrt{N^2-4}$.
This suggests that the surplus  specified by $\L^- = f$  should be identified with the $(0,f)$  primary in the CFT.
We will now generalize this observation to general surplus backgrounds by analyzing in detail the symmetries
of the surplus backgrounds. Comparing these with the large $c$ behavior of null states in the dual CFT  we will show that the surplus  specified by a general $\L^-$  should be identified with the $(0,\L^-)$  primary in the CFT.

A key point is   that the surplus backgrounds possess a large group of classical symmetries: each of them is invariant under an $sl(N)$ algebra. This easy to see from the observation made at the end of section \ref{secsurplus} that surplus backgrounds are globally pure gauge: they are gauge-equivalent to the trivial connection $A=0$ which is  invariant under all constant $sl(N)$ gauge transformations. Gauge transforming back to the highest weight gauge (\ref{hwgauge}) one finds
nonconstant gauge parameters which form an  $sl(N)$ and can be expressed as linear combinations of the asymptotic symmetry parameters $\z^s_m$ introduced above. We  denote
 the symmetry group of the surplus $a_{\L^-}$ by $sl(N)_{\L^-}$, and we propose that the $sl(N)_{\L^-}$ generators  agree
precisely with the twisted wedge modes in the CFT, discussed in section \ref{sectwisted}, which leave the CFT primary $|0,\L^-\rangle$ invariant in the large $c$ limit.
 In particular, the $N-1$ symmetry generators corresponding to the simple roots   of $sl(N)_{\L^-}$ correspond precisely  to the large $c$ limit of the operators which produce the independent  null states when acting on the primary $|0,\L^-\rangle$ in the CFT (denoted by $F_j$ in section \ref{sectwisted}).

We will again show this in detail for the cases $N=2$ and $N=3$. For $N=2$, one finds that the surplus specified by $\L^-$ is left invariant by the gauge transformations
\bea
e &=& {i \over r^-_1 +1} \z^2_{  r^-_1 +1 }\nonu
f &=& {i \over r^-_1 +1} \z^2_{- (  r^-_1 +1 )}\nonu
h &=&  -{2 \over r^-_1 +1} \z^2_{0} \label{sl2bulk}
\eea
These correspond precisely to the twisted wedge modes in the CFT (\ref{twistedsl2cft}), with $f$ corresponding to the operator that produces a null state when acting on the
$|0,\L^-\rangle$ primary.

For $N=3$, one finds that a general surplus background is invariant under the gauge transformations
\bea
e_1&=& {1 \over  \caln \sqrt{r^-_1-r^-_2+1}}\left( \sqrt{5\over 2}  \z^3_{r^-_1 - r^-_2+1}- {i \over 6} (r^-_1+r^-_2+3)
  \z^2_{r^-_1 - r^-_2+1}\right)\nonu
f_1&=&  {1 \over  \caln \sqrt{r^-_1-r^-_2+1}}\left( \sqrt{5\over 2}  \z^3_{- (r^-_1 - r^-_2+1)}- {i \over 6} (r^-_1+r^-_2+3)
  \z^2_{- (r^-_1 - r^-_2+1)}\right)\nonu
e_2 &=&  {1 \over  \caln \sqrt{r^-_2+1}}\left(\sqrt{5\over 2}  \z^3_{r^-_2+1} +{i\over 6} (2 r^-_1-r^-_2+3)  \z^2_{r^-_2+1}\right)\nonu
f_2 &=& {1 \over  \caln \sqrt{r^-_2+1}}\left(\sqrt{5\over 2}  \z^3_{-(r^-_2+1)} +{i\over 6} (2 r^-_1-r^-_2+3)  \z^2_{- (r^-_2+1)}\right)\nonu
h_1 &=&  {1 \over \caln^2} \left( {1\over 3} \left((r^-_1)^2-4 r^-_1 r^-_2+(r^-_2-6)
   r^-_2-3\right)\z^2_0 - i \sqrt{5 \over 2} (r^-_1+r^-_2+3)\z^3_0\right)\nonu
h_2 &=& {1 \over \caln^2} \left( {1\over 3}   \left(2 r^-_1 r^-_2-2 r^-_1
   (r^-_1+3)+(r^-_2)^2+6 r^-_2-3\right)\z^2_0 + i \sqrt{5 \over 2}  (2
   r^-_1-r^-_2+3) \z^3_0 \right)\nonu \label{sl3bulk}
\eea
where $\caln = \sqrt{(r^-_2+1)(r^-_1+2) (r^-_1-r^-_2+1)}$.
These again correspond precisely to the twisted wedge modes in the CFT (\ref{twistedsl3cft}), where the simple roots $f_1$  and $f_2$ correspond to the operators producing the independent null states at levels
$r^-_2+1$ and $r^-_1 - r^-_2+1$ respectively.

For general $N$, it is rather involved to compute the precise combination of generators that produces a symmetry,  but we can easily compute the $z$-frequencies of the symmetry parameters. These can then be compared to the energies of the null
 descendants in the dual CFT. Classical symmetries correspond to to solutions of the equation
\be
\d_\z a = \z' + [a, \z] = 0. \label{trivial}
\ee
Recall that for the  surplus backgrounds $a$ is constant and diagonalizeable, say by a matrix $M$:
\be
a_{\L^-} = M^{-1} \l M, \qquad \l =- i {\rm diag} (n^-_1, \ldots , n^-_N), \qquad n^-_i = \L^-_i  + {N+1\over 2} - i.\label{diaggauge}
\ee
The solutions to (\ref{trivial}) are as follows. There are $N-1$ constant solutions of the form $\z = M^{-1} \m M$ where $\m$ is diagonal
and traceless. Furthermore we have $N(N-1)$ nonconstant solutions
\be
\z_{ij} = e^{i (n^-_j-n^-_i) z}  M^{-1} e_{ji} M \qquad i \neq j.
\ee
where $e_{ij}$ are the standard basis of $N\times N$ matrices: $(e_{ij})_{kl} = \d_{ik} \d_{jl}$.
From this discussion we find symmetries with negative mode numbers at levels
\be
 n^-_i - n^-_j = \L^-_i - \L^-_j+ j - i \qquad i < j .\label{symmlevels}
\ee
For $j = i+1$, we find precisely the levels of the $N-1$ independent null vectors  in the $(0, \L^-)$ Verma module in the CFT (\ref{nulllevels}). These correspond to the
 simple roots of the $sl(N)_{\L^-}$ twisted wedge algebra.  The remaining  negative mode number symmetries \ref{symmlevels} correspond to the remaining positive roots of the twisted wedge algebra, while the positive and zero mode number symmetries correspond to negative roots and Cartan elements respectively.

To summarize this section, we have argued that the surplus solution specified by $\L^-$ is to be identified with the CFT primary $\ket{0, \L^-}$. Each
surplus  possesses an $sl(N)$ algebra of symmetries which is to be identified with the twisted wedge algebra $sl(N)_{\L^-}$ in the CFT. Note that we have focused again on the holomorphic sector; a similar derivation applies to the $\bar A$ gauge fields and  shows that each surplus has an additional antiholomorphic
$\overline{sl(N)}$ symmetry, which is to be identified with the antiholomorphic twisted wedge $\overline{sl(N)}_{\L^-}$ algebra in the CFT.

\subsection{Matter fluctuations around surpluses}\label{secscalarsurpl}

We will now study the fluctuations of the matter field $C$ around the surplus backgrounds. We will find that the solutions
 transform in the fundamental representation of the $sl(N)_{\L^-}$ symmetry of the surplus background parameterized by $\L^-$.
In parallel with the vacuum case $\L^-=0$, this representation is comprised of the primary $|f, \L^-\rangle$ and a particular subset of descendants obtained by acting with the wedge modes of the $sl(N)_{\L^-}$ with negative mode index.

Again, we choose to work in the  $r$-independent gauge introduced in section \ref{secsymms}, where the connection is simply $A = a_{\L^-} dz$. In this gauge the matter equations are
\bea
\pa_z \tilde C &=& - a_{\L^-} \tilde C\nonu
\pa_{\bar z} \tilde C &=&  - \tilde C \bar a_{\L^-}. \label{eqsrhoindepgauge}
\eea
To compute the energy and spin 3 charge of a matter solution we should compute its infinitesimal gauge variation with  parameter $\z^2_0$ and $\z^3_0$ respectively. Using  (\ref{gaugepars})
these are given by
\bea
\z^2_0 &=&  a_{\L^-}\nonu
\z^3_0 &=& 4 q \left( (a_{\L^-})^2 - {\tr (a_{\L^-})^2\over N} V^1_0 \right)
\eea
Using the equations of motion (\ref{eqsrhoindepgauge}), these  can be converted to differential operators acting on the trace part $\Phi$:
\bea
i \d_{\z^2_0} &\Leftrightarrow &i \pa_z \equiv l_0\nonu
i \d_{\z^3_0} &\Leftrightarrow & - i\sqrt{2 \over N^2-4} \left(\pa_z^2 + {(n^-)^2 \over N}\right) \equiv w_0.\label{0modediff}
\eea

The explicit solutions can be constructed in the same manner as the fluctuations around AdS.  We first find the solutions in a gauge where the gauge field is diagonal  and then transform back to the highest weight gauge (\ref{hwgauge}) with connection $a_{\L^-}$.
Taking the trace, this leads to $N^2$ solutions for the physical scalar which are, up to an inconsequential normalization constant, given simply by
\be
\Phi_{ij} \sim e^{i (n^-_i z - n^-_j \bar z)}
\ee
Since the eigenvalues $\{ n^-_i\}$ are distinct, these  solutions are linearly independent. Note that complex conjugation acts by exchanging $i$ and $j$, i.e.
$\overline{\Phi_{ij}} = \Phi_{ji}$.

Now let's compare the charges of the scalar solutions with those of CFT states.
The energies and spin 3 charges of these solutions can be read off by applying the differential operators (\ref{0modediff}):
\bea
 l_0 \cdot\Phi_{ij} &=&  - n^-_i \Phi_{ij}\nonu
 w_0 \cdot\Phi_{ij} &=& i\sqrt{2 \over N^2-4} \left( (n^-_i)^2 - {(n^-)^2 \over N} \right) \Phi_{ij}
 \eea
As was the case in the global AdS background, the state of lowest (left- and right-moving) energy is  $\Phi_{11}$. Comparing with (\ref{fLambdaweight}), (\ref{fLambdaspin3}) we see that it has the correct charges to be identified
with the primary $|f, \L^-\rangle$ in the CFT.
We propose that the other solutions $\Phi_{ij} $ are to be identified with a subset of descendants of the $|f, \L^-\rangle$, namely
 those obtained by acting with  left- and right-moving twisted wedge modes. If this proposal is true, the $N^2$ scalar solutions should
 transform as an $(N,N)$ representation of the $sl(N)_{\L^-}\times \overline{sl(N)}_{\L^-} $ symmetry of the background.
We will now verify this property for the cases  $N=2$ and $N=3$.

For $N=2$, there are four independent  solutions $\Phi_{ij}, \ i,j = 1,2$.
We would like to compute how they transform under the holomorphic $sl(2)_{\L^-}$ gauge transformations (\ref{sl2bulk}) (and their antiholomorphic cousins
$\overline{sl(2)}_{\L^-}$)  which leave the surplus background invariant. To do this, we act with the gauge transformations  (\ref{sl2bulk}) on the solutions $\Phi_{ij}$ in matrix form using the infinitesimal form of (\ref{gauget}), and then take the trace to find the action on $\Phi_{ij}$. Doing this one finds that the  $\Phi_{1j}$ are highest weight
states, $\d_e \Phi_{1j}=0$ while the  $\Phi_{2j}$ are lowest weight states, $\d_f \Phi_{2j}=0$. The solutions transform into each other
under $sl(2)_{\L^-}$ as follows:
\begin{center}
\begin{tikzpicture}
\matrix(m)[matrix of math nodes, row sep=2em, column sep=1.5em]
{\Phi_{1j} \\ \Phi_{2j} \\};
\path[->](m-1-1) edge [bend left] node [right] {$\delta_f$} (m-2-1);
\path[->](m-2-1) edge [bend left] node [left] {$\delta_e$} (m-1-1);
\end{tikzpicture}
\end{center}
The right-moving symmetries $\overline{sl(2)}_{\L^-}$ act in a similar manner on the on the column index $j$ of $\Phi_{ij}$.
In other words, the solutions $\Phi_{ij}$ fill out a $(2,2)$ representation under the $sl(2)_{\L^-}\times \overline{sl(2)}_{\L^-} $ symmetry
of the background,  just like their CFT duals in our proposed dictionary.

We can apply  the same method to analyze the transformation of the scalar solutions in the $N=3$ case under the left-moving $sl(3)_{\L^-}$ symmetries (\ref{sl3bulk}). The solutions $\Phi_{1j}$ are highest weight
states ($\d_{e_1} \Phi_{1j}=\d_{e_2} \Phi_{1j}=0$), while the solutions $\Phi_{3j}$ are lowest ($\d_{f_1} \Phi_{1j}=\d_{f_2} \Phi_{3j}=0$) weight, and the solutions transform into each other as
\begin{center}
\begin{tikzpicture}
\matrix(m)[matrix of math nodes, row sep=2em, column sep=1.5em]
{\Phi_{1j} \\ \Phi_{2j} \\ \Phi_{3j} \\};
\path[->](m-1-1) edge [bend left] node [right] {$\delta_{f_1}$} (m-2-1);
\path[->](m-2-1) edge [bend left] node [left] {$\delta_{e_1}$} (m-1-1);
\path[->](m-2-1) edge [bend left] node [right] {$\delta_{f_2}$} (m-3-1);
\path[->](m-3-1) edge [bend left] node [left] {$\delta_{e_2}$} (m-2-1);
\end{tikzpicture}
\end{center}
Combining with the similar result for the right-moving symmetries, we see that the scalar solutions indeed form a $(3,3)$ representation under the $sl(3)_{\L^-}\times \overline{sl(3)}_{\L^-} $ symmetry
of the background.

\subsection{Multiparticle states}\label{secmultipart}
So far we have identified the CFT primary $(0, \L^- )$ with the $a_{\L^-}$ surplus background and $(f, \L^- )$ with a single particle scalar excitation in this surplus background. We will now argue that the remaining CFT primaries $(\L^+, \L^- )$ correspond to specific  multiparticle excitations of the scalar field. We will do so by identifying their transformations under
the $sl(N)_{\L^-} \times \overline{sl(N)}_{\L^-}$ symmetry which we have argued organizes the spectrum on both sides of the duality.

We found that the single scalar excitation around the background transforms under $sl(N)_{\L^-} \times \overline{sl(N)}_{\L^-}$ as
\be
(\ydiagram{1},\ydiagram{1})
\ee
where $\ydiagram{1} \equiv f$, the fundamental representation.  Given the Bose statistics of the quantized field, the two particle-states transform under the symmetric square of this,
\be
\odot^2 (\ydiagram{1},\ydiagram{1}) = (\ydiagram{2},\ydiagram{2}) \oplus \left(\ydiagram{1,1},\ydiagram{1,1}\right)
\ee
Similarly, the three-particle states transform under
\be
\odot^3 (\ydiagram{1},\ydiagram{1}) = (\ydiagram{3},\ydiagram{3}) \oplus \left(\ydiagram{2,1},\ydiagram{2,1}\right) \oplus \left(\ydiagram{1,1,1},\ydiagram{1,1,1}\right)
\ee
In general, the decomposition of multiparticle states under $sl(N) \times sl(N)$ is given by
\be
\odot^k (\ydiagram{1},\ydiagram{1}) = \sum_{|\lambda|=k} (\lambda, \lambda) \label{boseproduct}
\ee
where the sum is over all Young diagrams of $k$ boxes.

We are thus led to the  conclusion that bulk $k$-particle states are associated to primaries with highest weight $(\Lambda^+, \Lambda^-)$ with respect to both the holomorphic $W_N$ and antiholomorphic $\bar W_N$ algebras, where $\Lambda^+$ has $k$ boxes.

It might be interesting to note that there is an analogous formula to (\ref{boseproduct}) for antisymmetric products (Fermi statistics). In that case, we have
\be
\wedge^k (\ydiagram{1},\ydiagram{1}) = \sum_{|\lambda|=k} (\lambda, \lambda^\prime)
\ee
where $\lambda^\prime$ is the dual representation to $\lambda$ (with transposed Young diagram). Here we would need to combine the chiral and antichiral $W_N$ representations in a non-diagonal way.

\section{Bulk partition function}\label{secpartbulk}
We are now ready to write down  the bulk 1-loop partition function of the Vasiliev theory at $\l = - N$. As we have argued, this partition function
will receive contributions from each classical surplus saddle point labeled by $\L^-$. The partition function around each of these saddle points
is a product of three contributions:  a classical contribution from the energy of the surplus, a 1-loop contributions from the
fluctuations of the higher spin gauge fields and another 1-loop  contribution from the spin-0 fluctuations around the surplus background. Schematically we write
\be
Z^{grav} = \sum_{\L^-} Z_{cl} (\L^-) Z_{1-loop}^{hs} (\L^-)  Z_{1-loop}^{scalar} (\L^-).
\ee

The classical contribution can be read off from the energy of the surplus solutions in (\ref{charges})
\be
 Z_{cl} (\L^-) =  |q|^{- {2 C_2(n^-) \over N(N^2-1)}c }.\label{Zclass}
 \ee

 The 1-loop contributions to the partition function count noninteracting single- and multiparticle states in the second-quantized theory around
 a given surplus background. Our strategy for counting multiparticle  states will be as follows: from our discussion in sections \ref{secsymms},\ref{secscalarsurpl} we know the single particle spectrum and hence the single particle
  partition function. The multiparticle partition function can be computed from the standard expression for free Bose particles:
  \be
  Z =  \exp \left[ \sum_{n=1}^\infty {Z_{1-part.}(q^n, \bar q^n) \over n} \right] .\label{multipart}
 \ee
 We will do this in turn for the  higher spin gauge fields and the scalar field.

\subsection{Counting boundary higher spin particles}

   Let us start with the simple example of pure gravity ($N=2$) around the global AdS background $\L^-=0$. We argued
   in section \ref{secsymms} that the single particle boundary gravitons  correspond to the linearized fluctuations $\d_{\z^2_{-n}}a_{AdS}$ for $n\geq2$ (recall that
   $\z^2_{-1}$ generates a symmetry and hence doesn't correspond to a boundary graviton).
 From this we find the single particle partition function
\be
Z_{1-part} (q,\bar q)  = ( q^2 + q^3 + q^4 + \ldots) ( \bar q^2 + \bar q^3 +\bar q^4 + \ldots)= {|q|^4 \over |1- q|^2}
\ee
and using (\ref{multipart}) we obtain for the multiparticle partition function
\be
Z =  \prod_{n=2}^\infty {1 \over | 1 - q^n|^2}.
\ee
This indeed agrees with the 1-loop gravity partition function in  AdS background computed using heat kernel methods  in \cite{Giombi:2008vd}.

We now return to the general case. The single particle states around a surplus background $a_{\L^-}$ are the linearized fluctuations $\d_{\z^s_{-n}}a_{\L^-}, n>0$ modulo those combinations which generate symmetries.   We saw in (\ref{symmlevels}) that the  surplus background has $N(N-1)/2$ negative mode  symmetry generators at
levels  $n^-_i - n^-_j$ with $i>j$.
Including also the contribution from the antiholomorphic sector, we find for the single particle partition function
\bea
Z^{hs}_{1-part} (\L^-) &=& \left((N-1) ( q + q^2 + q^3 + \ldots )  - \sum_{1 \leq i < j \leq N} q^{n^-_i - n^-_j}\right) \times \nonu
&& \left((N-1) ( \bar q +\bar  q^2 +\bar  q^3 + \ldots )  - \sum_{1 \leq i < j \leq N} \bar q^{n^-_i - n^-_j}\right)
\eea
and for the multiparticle partition function, using (\ref{multipart}),
\be
Z_{1-loop}^{hs} (\L^-)  ={ |q|^{N-1 \over 12} \prod_{1 \leq i < j \leq N} | 1 - q^{n^-_i - n^-_j} |^2 \over |\h|^{2(N-1)}}.\label{Zhs}
\ee
As a check, we work this out for the global  AdS background using (\ref{AdSns}) and find
\be
Z_{1-loop}^{hs} (0)= \prod_{s = 2}^N \prod_{n=s}^\infty {1 \over |1 - q^n|^2}.
\ee
This indeed reproduces the the known  result for the 1-loop determinant for linearized higher spin fields of spins $s=2,3,\ldots N$ in global AdS \cite{Gaberdiel:2010ar}.

\subsection{Counting scalar particles}

Now let's consider  the 1-loop scalar partition function. We saw that there are $N^2$ single particle states whose energies $\D h, \D \bar h$
are determined by the eigenvalues of the surplus connection and given by
  $- n^-_i  $. Hence the single particle contribution to the partition function is
\be
Z^{scalar}_{1-part.}(q,\bar q) = (q^{-n^-_1} + \ldots +q^{-n^-_N})(\bar q^{-n^-_1} + \ldots +\bar q^{-n^-_N})
\ee
To compare with the CFT side, it's useful to rewrite this using the techniques from \cite{Gaberdiel:2011zw}. The single particle contribution
can be written in terms of $u(N)$ characters:
\be
Z^{scalar}_{1-part.}(q,\bar q) = \tr_f U \tr_f \bar U.
\ee
where $U = e^{-2 \p i \t {\rm diag} (n^-)}$.
To include multiparticle states we can again use (\ref{multipart}), leading to
\bea
Z^{scalar}_{1-loop}(\L^-)
&=& \sum_{\L^+}  \tr_{\L^+} U \tr_{\L^+} \bar U \nonu
&=& \sum_{\L^+} | \chi_{\L^+} ( - 2 \p i \t n^- ) )|^2. \label{Zscal}
\eea
where, in the second step, we have made use of an identity derived   in  section 3 of \cite{Gaberdiel:2011zw} which reflects the representation-theoretical statement (\ref{boseproduct}).

Multiplying the contributions (\ref{Zclass}), (\ref{Zhs}), (\ref{Zscal}), we see that despite being somewhat crude in our methods,  we  reproduce the large $c$  CFT partition function (\ref{CFTZlargec})
almost exactly. The bulk and CFT results differ by  an overall factor $(q\bar q)^{{C_2  (n^- )\over N(N+1)} + (\L^- , n^- ) -{N-1 \over 24} }$, which we see from (\ref{O0weight}) represents precisely  the 1-loop correction $h^{(0)}_{(0,\Lambda^-)}$ to the energy
of the $(0, \L^- )$ primary. This comes from backreaction of $\calo(c^0)$ quanta; we worked on a fixed background, so its absence is unsurprising. It would be interesting to reproduce this correction from a careful treatment of the path integral over the higher spin fields.

\section{Discussion}

In this work we studied a specific  corner of the duality between conformal field theories with extended  $W$-symmetry and higher spin gravity.
In particular we compared the spectrum of $W_N$ conformal field theories at large value of the central charge to that  of semiclassical
 Vasiliev theory at  $\l = -N$.
 Despite this being a non-unitary corner of the duality, it is also a very tractable one due the fact that the Vasiliev
 theory reduces to an  $sl(N)$ Chern-Simons theory, and the linearized matter couplings are fixed by this symmetry.
 This tractability allowed us to explore  some of the finer points of the  AdS/CFT dictionary.

 One of these  points is the precise identification of conical surplus solutions with CFT primaries. By a careful comparison of the classical symmetries in the bulk and the symmetries which emerge in the  large $c$ limit of the CFT, we argued that the surpluses are to be identified
 with the $(0, \L^-)$ states.

 Another point we addressed is the content of the equations for the matter field in the $\l = -N$ Vasiliev theory.  We showed that its
 equation of motion describes a discrete set of states, which transform in a non-unitary finite dimensional representation of the symmetry group of the background. By studying the  symmetry properties of scalar solutions in the surplus backgrounds we argued that  the single particle states of the matter field
 are to be identified with the primary  $|f, \L^-\rangle$ and certain descendants.
 We also showed  that more general primaries $|\L^+, \L^- \rangle$ are correspond to  multiparticle excitations of the matter field around a surplus background.

 This implies for example that the states of the form $|\L, \L \rangle$ are dual to a particular multiparticle state in a surplus background.
 These are the  states which become light in the 't Hooft limit of the $W_N$ minimal models, and have so far been poorly understood from the bulk point of view. It would be of great interest to understand how our dictionary for these states in our semiclassical regime carries over to the 't Hooft regime of the duality. Can it come to bear on the puzzles of \cite{Banerjee:2012aj}, for example?

The results on linearized scalar matter apply generally in the context of the bulk theory. An interesting open question is whether this matter field, despite the  non-unitary constraints, can be used to understand this duality at the level of interactions and physical processes. For instance, one might ask whether we can probe higher spin backgrounds a la \cite{Kraus:2012uf}, for instance by placing a scalar in the higher spin  black hole backgrounds of \cite{Gutperle:2011kf}. Along these lines, similar computations -- the comparison of correlation functions, studying the contribution of higher  spin black hole solutions to
  the thermal partition function, and so on -- would help establish whether there exists a bona fide duality in the non-unitary regime.

  Our analysis has also revealed  an interesting structure in the large $c$ limit  of $W_N$ CFTs, namely the existence, in each nonperturbative sector,
  of linear combinations of the $W_N$ generators, the `twisted wedge'
  modes, which form an $sl(N)$ algebra at large $c$ which organizes the spectrum in that sector.
We verified these statements only of the simplest cases $N=2$ and $N=3$, but the dual picture predicts them to generalize to all $N$ as indicated in Section 3.
It would therefore be interesting to have a CFT derivation of these properties.

It would also be of interest to give a more rigorous  path-integral derivation of the one-loop higher spin partition function around a surplus
background, generalizing the derivation of \cite{Gaberdiel:2010ar} in AdS. This would be a main step toward a bulk derivation of  the one loop shift $h^{(0)}_{(0,\L^-)}$ of the energy of the surplus backgrounds  which was not captured by our current computation.

\section*{Acknowledgements}
We would like to thank M. Ammon, A. Campoleoni, A. Castro, F. Denef, M. Gutperle, P. Kraus, M. Rangamani, X. Yin and especially R. Gopakumar for useful discussions.

 The work of J.R. has been supported  in part by the Czech Science Foundation  grant GACR P203/11/1388 and in part by the EURYI grant GACR  EYI/07/E010 from EUROHORC and ESF.  The work of T.P. has been supported  in part by the EURYI grant GACR  EYI/07/E010 from EUROHORC and ESF. E.P. has received funding from the European Research Council under the European Union's Seventh Framework Programme (FP7/2007-2013), ERC Grant agreement STG 279943, “Strongly Coupled Systems”.

\begin{appendix}

\section{Low spin commutators of $W_N$}\label{appWn}
The $W_N$ commutation relations for some of the low spin modes are
\begin{align}
{}[L_m, L_n] &= (m-n)L_{m+n} + \frac{c}{12}m(m^2-1)\delta_{m,-n}\nonu
{}[L_m, W^s_n] &= \left((s-1)m-n\right)W^s_{m+n}~, ~~ s>2\nonu
{}[W^3_m, W^3_n] &= 2(m-n)W^4_{m+n}+  \frac{1}{30}(m-n)(2m^2+2n^2-mn-8)L_{m+n} \nonu & \ \ \ \
+\frac{16}{(5c+22)}(m-n)\Lambda^{(4)}_{m+n}
+ \frac{c}{3 \cdot 5!}m(m^2-1)(m^2-4)\delta_{m,-n}\label{wcomm}
\end{align}
where
\bea
\Lambda^{(4)}_n & = & \sum_{p} : L_{n-p} L_p :  + \tfrac{1}{5} x_n L_n
 \nonu
 x_{2l} &=& (l+1)(1-l) \ , \qquad   x_{2l-1} = (l+1) (2-l) .
 \eea
\section{Details on scalar field equations}
\subsection{Constraints on scalar solutions}\label{appconstr}
Here we derive some constraints on the physical matter field $\Phi$ from the equations of motion (\ref{mattereqs}).
With $a_{\zb}=0$, we find the following constraints:
\bea
\prod_{j = 1}^N (\pa_r + N+1 - 2 j ) \Phi &=& 0\label{constr1app}\\
\prod_{j = 1}^N (\pa_z + D_j ) \Phi &=& 0\label{constr2}\\
\prod_{j = 1}^N (\pa_{\bar z} + \bar D_j ) \Phi &=& 0.\label{constr3}
\eea
Here we have assumed that $a$ is diagonalizeable with eigenvalues $D_i$.

To show the first property, note that (\ref{rhoeq}) implies for the diagonal elements of $C$:
$$ (\pa_r + N+1 - 2 j) C_{jj} \qquad {\rm (no\ sum).}$$ Hence the trace $\Phi = \sum C_{jj}$ must be annihilated by the
product of the differential operators, leading to (\ref{constr1app}).

To prove the second property, note that $A_z = (M b)^{-1} D (M b)$, where $M$ is the matrix that diagonalizes $a$ and
$D= {\rm diag} (D_1,\ldots, D_N )$. Plugging into (\ref{zeq}) we see that $\tilde C = (M b) C (M b)^{-1}$ satisfies
\be
 (\pa_z + D_j ) \tilde C_{jj} \qquad {\rm (no\ sum).}\ee
Again the trace must be annihilated by the product of the differential operators, leading to (\ref{constr2}).
The third property is proved analogously, by considering $\tilde {\tilde C} = (b M^\dagger)^{-1} C (b M^\dagger)$.

\subsection{$sl(2)$ action on scalar solutions in global AdS}\label{appsl2scalar}
We first write the $z$ and $\bar{z}$ equations of motion for the matter field $C$ in the AdS background,
\begin{eqnarray}
\partial_z C & = & -\frac{1}{2} \left( e^{r} L_+ C + e^{-r} L_- C \right) \nonu
\partial_{\bar{z}} C & = & \frac{1}{2} \left( e^{-r} C L_+ + e^{r} C L_- \right)
\end{eqnarray}
Taking the trace, this can be solved for $\tr L_{\pm} C$,
\begin{equation}
\tr \left(L_{\pm} C \right) = \mp \tr \frac{e^{\pm r} \partial_z C + e^{\mp r} \partial_{\bar{z}} C}{\sinh 2r}
\end{equation}
Using this, we can write
\begin{eqnarray}
\delta_{l_{\pm}} \Phi & \equiv & {1\over N}\tr \left(-\Lambda(L_{\pm}) C \right) =  {1\over N} e^{\pm i z} \tr \left( \mp L_0 C - \frac{i}{2} e^{r} L_+ C + \frac{i}{2} e^{-r} L_- C \right) \nonu
& = & e^{\pm i z} \left( \pm \frac{1}{2} \partial_{r} + \frac{i \cosh 2r}{\sinh 2r} \partial_z + \frac{i}{\sinh 2r} \partial_{\bar{z}} \right) \Phi
\end{eqnarray}
Hence the $sl(2)$ transformations act on $\Phi$ as the standard Killing vectors of AdS:
\bea
l_0 &=& i \pa_z \nonu
l_{\pm 1} &=& i e^{\pm i z}\left( \coth 2 r \pa_z + {1 \over \sinh 2r} \pa_{\bar z} \mp {i \over 2} \pa_r\right)\label{KVsapp}
\eea
 The antiholomorphic generators $\bar l_m$ are obtained by complex conjugation.
\subsection{A scalar in Poincar\'e AdS}
We would like to put a scalar in Poincar\'e AdS, with connection
\be
a = V^2_1 dz
\ee
Let us fix $\l=-2$ for simplicity. Using the `brute force'approach of expanding $C$ in components and solving the system of differential equations, one finds, as in global AdS, four independent solutions:
\be
\Phi = \a_1e^r(e^{-2r} + z\zb) + \a_2 e^r z + \a_3 e^r \zb + \a_4 e^r
\ee
One of them is a highest weight state,
\be\label{poinprim}
\Phi_{11} = e^r(e^{-2r} + z\zb)
\ee
and the others are descendants obtained by acting with $\pa_z,\pa_{\zb}$:
\be
\Phi = \Phi_{11} + \a_1 \pa_{\zb}\Phi_{11} + \a_2 \pa_z \Phi_{11} + \a_3 \pa_{\zb}\pa_z\Phi_{11}
\ee
In Poincar\'e coordinates (in contrast to global coordinates), this is just the action of lowering operators: that is, $l_- \sim \pa_z, \bar l_- \sim \pa_{\zb}$, with $l_1,\bar l_-$ each living in $sl(2)$. So we again find a (2+2)-dimensional representation of $sl(2) \times \overline{sl(2)}$.

Taking into account a Lorentzian vs. Euclidean signature convention and coordinate differences, these results are precisely those in section 4.3 of \cite{Balasubramanian:1998sn}. For instance, the primary (\ref{poinprim}) is given in their equation (79) where we recall $h=-1/2$ for our scalar field. (Note that $r_{there} = e^{-r}_{here}$ and $z\zb_{there} = -z\zb_{here}$.)

\section{Chevalley basis for the $sl(3)$ Lie algebra}\label{appsl3}
In this appendix we summarize our choice of $sl(3)$ basis and their commutation relations. In the fundamental representation, we have the following generators (in the Chevalley basis)
\begin{eqnarray}
E_1 = \begin{pmatrix} 0 & 1 & 0 \\ 0 & 0 & 0 \\ 0 & 0 & 0 \end{pmatrix} \quad & \quad F_1 = \begin{pmatrix} 0 & 0 & 0 \\ 1 & 0 & 0 \\ 0 & 0 & 0 \end{pmatrix} \quad & \quad H_1 = \begin{pmatrix} 1 & 0 & 0 \\ 0 & -1 & 0 \\ 0 & 0 & 0 \end{pmatrix} \nonu
E_2 = \begin{pmatrix} 0 & 0 & 0 \\ 0 & 0 & 1 \\ 0 & 0 & 0 \end{pmatrix} \quad & \quad F_2 = \begin{pmatrix} 0 & 0 & 0 \\ 0 & 0 & 0 \\ 0 & 1 & 0 \end{pmatrix} \quad & \quad H_2 = \begin{pmatrix} 0 & 0 & 0 \\ 0 & 1 & 0 \\ 0 & 0 & -1 \end{pmatrix}
\end{eqnarray}
satisfying commutation relations
\begin{center}
\begin{tabular}{llll}
$\left[H_1, E_1\right] = 2E_1$ & $\left[H_2, E_1\right] = -E_1$ & $\left[H_1, E_2\right] = -E_2 $ & $\left[H_2, E_2\right] = 2E_2$ \\
$\left[H_1, F_1\right] = -2F_1$ & $\left[H_2, F_1\right] = F_1$ & $\left[H_1, F_2\right] = F_2 $ & $\left[H_2, F_2\right] = -2F_2$ \\
$\left[E_1, F_1\right] = H_1$ & $\left[E_2, F_2\right] = H_2$ & $\left[E_1, F_2\right] = 0$ & $\left[E_2, F_1\right] = 0$ \\
$\left[E_1, \left[E_1, E_2\right]\right] = 0$ & $\left[E_2, \left[E_1, E_2\right]\right] = 0$ & $\left[F_1, \left[F_1, F_2\right]\right] = 0$ & $\left[F_2, \left[F_1, F_2\right]\right] = 0$ \\
$\left[H_1, H_2\right] = 0$ \\
\end{tabular}
\end{center}
Of course, not all of these are independent, as we can see using the Jacobi identities. With this choice of generators, one can check that the action of the Cartan subalgebra generators $H_j$ on the highest weight state of the representation $\Lambda$ is
\be
H_j \ket{\Lambda}  =  d_j \ket{\Lambda}\label{hwsl3}
\ee
where $d_j \equiv r_j - r_{j+1}$ are the Dynkin labels of $\L$.

\section{Null states}\label{appnull}

Here we collect some explicit expressions for the null states of various $(\L^+, \L^-)$ representations.

Of all of the null vectors, only $N-1$  vectors are `independent' in the sense that all the others appear as descendants of those.  The independent null vectors appear at levels
\be
(\L^+_{j}-\L^+_{j+1} +1)(\L^-_{j}-\L^-_{j+1} +1) \qquad j = 1, \ldots , N-1.
\ee

For the vacuum representation $\L^+ = \L^- = 0$, the $N-1$ independent null states all appear at level one and hence must be  given by
\be L_{-1} |0\rangle,\ W^{3}_{-1} |0\rangle,\ \ldots , W^{N-1}_{-1} |0\rangle.\ee
This implies  the invariance of the vacuum under the wedge modes (\ref{cftvacinv}).

The null vectors in the $(f,0)$ and $(0,f)$ representations were worked out for  general $N$ in \cite{Gaberdiel:2012ku}.
Their charges are, from (\ref{weight}), (\ref{spin3charge})
\bea
h_{(f,0)} &=& {(N-1)(2N + k + 1)\over 2N(N+k)}, \ \  w_{3,(f,0)} = - {(N-1)(2N + k + 1)\over 3 \sqrt{2}  N(N+k)}\sqrt{(N-2) (3N + 2k +2)\over N(N+ 2 k)}\nonu
h_{(0,f)} &=& {(N-1) k\over 2 N (N+k+1)},  \ \  \ \ \ \ \ \ \ w_{3,(0,f)} = {(N-1) k\over 3 \sqrt{2} N (N+k+1)} \sqrt{(N-2)(N+ 2k)\over N (3N + 2k +2)}.
\eea
There are $N-2$ independent null vectors at level one, given by
\be
\caln_{1,\chi}^s = \left( W^s_{-1} - {s w_s \over 2 h} L_{-1} \right) |\chi\rangle \qquad s = 3,\ldots , N,
\ee
and another independent null vector at level 2:
\be
\caln_{2,\chi} = \left( W^3_{-2} - {3 w_3 (2 h + c)\over h( 16h^2 + c (2h+1) - 10h )}  L_{-1}^2 -   {24 w_3 (h-1)\over ( 16h^2 + c (2h+1) - 10h )}   L_{-2} \right) |\chi\rangle.
\ee

The level one null vector $\caln_1^3$ and  level two null vector $\caln_2$ of the $(f,0)$ and $(0,f)$ representations become, in the large $c$ limit:
\bea
\caln_{1,(0,f)}^3 = \left( W^{(3)}_{-1} - {\sqrt{4 - N^2}\over \sqrt{2}N} L_{-1}  + \calo(1/c) \right) |0,f\rangle\nonu
\caln_{2,(0,f)} = \left( W^{(3)}_{-2} - {2 \sqrt{2} \over N \sqrt{4-N^2}}L_{-2}  + \calo(1/c)\right) |0,f\rangle\nonu
\caln_{1,(f,0)}^3 = \left( W^{(3)}_{-1} + \sqrt{2-N\over 2(2+N)} L_{-1} + \calo(1/c) \right) |f,0\rangle\nonu
\caln_{2,(f,0)} = \left( W^{(3)}_{-2} + \sqrt{2 \over  4-N^2} L_{-1}^2  + \calo(1/c)\right) |f,0\rangle \label{fnullvects}
\eea

In principle, knowledge of the spectrum of null states is sufficient to write the desired character. For simple non-vacuum representations like $(f,0)$ and $(0,f)$, for example,
the reasoning around (\ref{polcorr}) also works, and the character is given simply by
\be\label{ch0f}
{\rm ch}_{(f,0)} = {\rm ch}_{(0,0)} \times {1-q^N\over 1-q}
\ee
This result can be understood upon knowing that the $(f,0)$ representation has $N-2$ independent null vectors at level one, one independent null vector at level two, and counting their descendants.

\subsection{Null states for $N=3$ in the large $c$ limit}
It is useful to know some explicit expressions for the null states in the large $c$ limit. For $N=3$ the results are summarized in table \ref{nulltab}.
\begin{table}[h]
\centering
\begin{tabular}{|c|c|c|c|}
\hline
rep & $h$, $w_3$ & level & large $c$ limit of null state \\
\hline
\hline
$(0,0)$ & $0$ & $1$ & $W_{-1} - \frac{i}{\sqrt{10}} L_{-1}\sim F_1$ \\
        & $0$ & $1$ & $W_{-1} + \frac{i}{\sqrt{10}} L_{-1}\sim F_2$ \\
\hline
$(f,0)$ & $-1$ & $2 \cdot 1$ & $\left(W_{-1} - \frac{i}{\sqrt{10}} L_{-1}\right)^2\sim (F_1)^2$ \\
        & $\frac{i}{3} \sqrt{\frac{2}{5}}$ & $1$ & $W_{-1} + \frac{i}{\sqrt{10}} L_{-1}\sim F_2$ \\
\hline
\hline
$(0,f)$ & $-\frac{c}{18} + \frac{16}{9}$ & $2$ & $W_{-2} -\frac{2i}{3} \sqrt{\frac{2}{5}} L_{-2}\sim F_1$ \\
        & $\frac{i}{81} \sqrt{\frac{5}{2}}c -\frac{112i}{81} \sqrt{\frac{2}{5}}$ & $1$ & $W_{-1} + \frac{i}{3} \sqrt{\frac{5}{2}} L_{-1}\sim F_2$ \\
\hline
$(f,f)$ & $-\frac{c}{18} + \frac{1}{9}$ & $2 \cdot 2$ & $\left(W_{-2} -\frac{2i}{3} \sqrt{\frac{2}{5}} L_{-2}\right)^2\sim (F_1)^2$ \\
        & $\frac{i}{81} \sqrt{\frac{5}{2}}c -\frac{13i}{81} \sqrt{\frac{2}{5}}$ & $1$ & $W_{-1} + \frac{i}{3} \sqrt{\frac{5}{2}} L_{-1}\sim F_2$ \\
\hline
$(\bar{f},f)$ & $-\frac{c}{18} + \frac{4}{9}$ & $2$ & $W_{-2} -\frac{2i}{3} \sqrt{\frac{2}{5}} L_{-2}\sim F_1$ \\
              & $\frac{i}{81} \sqrt{\frac{5}{2}}c -\frac{26\sqrt{10}i}{81}$ & $2 \cdot 1$ & $\left(W_{-1} + \frac{i}{3} \sqrt{\frac{5}{2}} L_{-1}\right)^2\sim (F_2)^2$ \\
\hline
\hline
$(0, \odot^2 f)$ & $-\frac{5c}{36} + \frac{89}{18}$ & 3 & $W_{-3} -\frac{i}{3} \sqrt{\frac{5}{2}} L_{-3}\sim F_1$ \\
                 & $\frac{7i}{162} \sqrt{\frac{5}{2}}c -\frac{919i}{81\sqrt{10}}$ & 1 & $W_{-1} +\frac{7i}{3\sqrt{10}} L_{-1}\sim F_2$ \\
\hline
\end{tabular}
\caption{Null states of some $N=3$ representations at large $c$. In the last column we indicated that the operators are proportional to powers of the twisted wedge modes defined in (\ref{twistedsl3cft}).}
\label{nulltab}
\end{table}
We obtained these expressions by finding the null states for arbitrary $c$ and taking the limit $c\to\infty$ of the resulting expressions. Note the linearity in the generators of the null states for the $(0,\L^-)$ representations.

\section{Structure of $N=3$ Verma modules}\label{appVerma}

In this appendix we argue how  the formula for the correction factor  (\ref{polcorr}) for the vacuum character due to null states  generalizes to  arbitrary $(\Lambda^+,\Lambda^-)$ representations.  In general, (\ref{polcorr})  is  replaced by a more complicated polynomial which encodes the structure of the $(\Lambda^+,\Lambda^-)$ Verma module. For the regime of interest (large and generic $c$)  this structure was worked out in \cite{Niedermaier:1991cu} which we will now review and motivate
using examples.

 It
 is useful to first rederive  (\ref{polcorr}) using a more involved method which however generalizes to general representations. This consists of
 finding all Verma submodules and the pattern of inclusions between these submodules. As a warmup example, we do this explicitly for the $(0,0)$ and $(f,0)$ representations  $N=3$.

\subsection{Vacuum module $(0,0)$}

We start from the highest weight vector
\be
\ket{h=0, w=0}
\ee
Both of its descendants at level $1$ are null primaries,
\be
\ket{h=1, w = \pm \sqrt{\frac{2(2-c)}{5c+22}}} \sim \left(\pm \sqrt{\frac{2-c}{2(5c+22)}} L_{-1} + W_{-1} \right) \ket{h=0,w=0}.
\ee
The corresponding two null Verma submodules intersect at level 3 of the $\ket{h=0, w=0}$ Verma module. The intersection of these is a sum of another two Verma submodules
\be
\ket{h=3, w = \pm \sqrt{\frac{2(98-c)}{5c+22}}}
\ee
Finally, these two Verma submodules intersect at level $4$ of $\ket{h=0, w=0}$ where there is the last null primary of $\ket{h=0,w=0}$ Verma module. Diagramatically, we have the following structure of Verma submodules inclusions:
\begin{center}
\begin{tikzpicture}
\matrix(m)[matrix of math nodes, row sep=3em, column sep=1.5em]
{& \ket{h=0, w=0} & \\
\ket{h=1,w=\sqrt{\frac{2(2-c)}{5c+22}}} & & \ket{h=1,w=-\sqrt{\frac{2(2-c)}{5c+22}}} \\
\ket{h=3,w=\sqrt{\frac{2(98-c)}{5c+22}}} & & \ket{h=3,w=-\sqrt{\frac{2(98-c)}{5c+22}}} \\
& \ket{h=4,w=0} & \\};
\path[->](m-1-2) edge (m-2-1);
\path[->](m-1-2) edge (m-2-3);
\path[->](m-2-1) edge (m-3-1);
\path[->](m-2-1) edge (m-3-3);
\path[->](m-2-3) edge (m-3-1);
\path[->](m-2-3) edge (m-3-3);
\path[->](m-3-1) edge (m-4-2);
\path[->](m-3-3) edge (m-4-2);
\end{tikzpicture}
\end{center}
This should be compared to the numerator in the character formula which is now
\be\label{e4}
(1-q)^2(1-q^2) = 1-2q+2q^3-q^4
\ee
We first remove two Verma submodules at level $1$. This, however, removed the level $3$ submodules twice so we should add them to compensate for this. Finally, we remove the Verma submodule at level $4$ (which was so far twice removed and twice added).

We note that while this approach is guaranteed to give the correct answer, the direct connection between this and the factorized form of the correction -- in which each term lies in one-to-one correspondence with generators of null states -- is not completely self-evident.

\subsection{$(f,0)$ module}

Let us consider the structure of Verma module with the highest weight $\ket{\Lambda^+ = f, \Lambda^-=0}$, still working in $N=3$. The structure of the submodules is similar to the case of vaccum module, but the Verma submodules appear at different levels. There are two submodules at level 1 and 2 of the highest weight Verma module. They intersect in sum of another two submodules at levels 4 and 5. Finally, there is a submodule at level 6 which is an intersection of these two submodules. The following diagram illustrates these embeddings:
\begin{center}
\begin{tikzpicture}
\matrix(m)[matrix of math nodes, row sep=3em, column sep=-9em]
{& \ket{h=\frac{k+7}{3(k+3)}, w=\frac{\sqrt{2}(k+7)(2k+11)}{9(k+3)\sqrt{3(2k+3)(2k+11)}}} & \\
\ket{h=\frac{k+7}{3(k+3)}+2, w=\frac{5\sqrt{2}(k+7)(4k+13)}{9(k+3)\sqrt{3(2k+3)(2k+11)}}} & & \ket{h=\frac{k+7}{3(k+3)}+1, w=-\frac{4\sqrt{2}(2k+5)(2k+11)}{9(k+3)\sqrt{3(2k+3)(2k+11)}}} \\
\ket{h=\frac{k+7}{3(k+3)}+5, w=\frac{\sqrt{2}(7k+25)(8k+29)}{9(k+3)\sqrt{3(2k+3)(2k+11)}}} & & \ket{h=\frac{k+7}{3(k+3)}+4, w=-\frac{5\sqrt{2}(2k+5)(7k+25)}{9(k+3)\sqrt{3(2k+3)(2k+11)}}} \\
& \ket{h=\frac{k+7}{3(k+3)}+6,w=\frac{\sqrt{2}(7k+25)(8k+29)}{9(k+3)\sqrt{3(2k+3)(2k+11)}}} & \\};
\path[->](m-1-2) edge (m-2-1);
\path[->](m-1-2) edge (m-2-3);
\path[->](m-2-1) edge (m-3-1);
\path[->](m-2-1) edge (m-3-3);
\path[->](m-2-3) edge (m-3-1);
\path[->](m-2-3) edge (m-3-3);
\path[->](m-3-1) edge (m-4-2);
\path[->](m-3-3) edge (m-4-2);
\end{tikzpicture}
\end{center}
Again, the character formula is a direct consequence of this embedding pattern. In fact, the character is equal to
\be
{\rm ch}_{(f,0)} = q^{h(f,0)-{c \over 24}} \frac{1-q-q^2+q^4+q^5-q^6}{\prod_{n=1}^\infty (1-q^n)^2}
\ee
which instructs us to take the Verma module corresponding to the highest weight, subtract the null submodules at level 1 and 2, add null submodules at level 4 and 5 (since they were subtacted twice) and finally subtract the submodule at level 6. This result factorizes and matches (\ref{ch0f}), given (\ref{e4}).

\subsection{$(\L^+, \L^- )$ modules}
The above examples show that the $(0,0)$ and $(f,0)$ representations  have the same structure of Verma module inclusions, with two null submodules at the highest level and $6$ submodules in total (including the Verma module itself). In fact, all $(\Lambda^+,\Lambda^-)$ representations for $N=3$ have this structure.
The underlying pattern is as follows. Turning now to general $N$, there will be in total $N!$ Verma submodules which are in one-to-one correspondence with the elements of the Weyl group of $sl(N)$. The null vector corresponding to the Weyl element $w$ is labeled by 
\be
(w \cdot \Lambda^+, \Lambda^-)
\ee
where
\be
w \cdot \Lambda \equiv w(\Lambda+\rho)-\rho
\ee
is the shifted Weyl reflection. The embedding pattern of submodules depends only on $N$ and not on the highest weights $(\Lambda^+,\Lambda^-)$. In fact, it is determined by so called strong Bruhat order of elements of the Weyl group.

The $N-1$ simple Weyl reflections give rise to  $N-1$  `independent' null vectors which have the property that all other null vectors are contained in their Verma modules.
 These independent null vectors appear at levels
\be
(\L^+_{j}-\L^+_{j+1}+1)(\L^-_{j}-\L^-_{j+1}+1) ~,\qquad j = 1, \ldots , N-1.\label{nulllevelsapp}
\ee
From the point of view of relations between $W_N$, these null states correspond precisely to $N-1$ independent additional relations between $W_N$ generators that hold in the  $(\Lambda^+,\Lambda^-)$ representation apart from the commutation relations themselves.

Factoring out the union of all the Verma submodules, we obtain the irreducible representation $(\Lambda^+, \Lambda^-)$.
 In terms of characters,  this procedure leads to equation (\ref{char}).
\end{appendix}

\end{document}